\documentclass[twoside]{article}
\usepackage{epsfig,latexsym}
\usepackage[smalltableaux]{ytableau}
\usepackage{float,lipsum,enumitem,graphicx,csquotes,tikz,makecell,setspace,pifont,moresize}

\usepackage{amsmath,bbm,bm,mathrsfs,amsthm,amssymb,dsfont,mathdots,stmaryrd}
\usepackage{hyperref}
\usepackage{qic}

\newcommand{\nc}{\newcommand}       \newcommand{\rnc}{\renewcommand}
\nc{\x}{\textnormal} \nc{\n}{\operatorname} \nc{\s}{\mathsf} \nc{\bb}{\mathbb}
\nc{\alp}{\alpha}  \nc{\bt}{\beta}      \nc{\gm}{\gamma}  \nc{\Gm}{\Gamma} \nc{\dt}{\delta}
\nc{\Dt}{\Delta}   \nc{\kp}{\kappa}     \nc{\sg}{\sigma}  \nc{\Sg}{\Sigma} \nc{\tht}{\theta}
\nc{\Tht}{\Theta}  \nc{\ld}{\lambda}    \nc{\Ld}{\Lambda} \nc{\om}{\omega} \nc{\Om}{\Omega}
\nc{\phv}{\varphi} \nc{\epsl}{\varepsilon}  \nc{\thv}{\vartheta}
\nc{\Cal}[1]{\mathcal{#1}} \nc{\fr}[1]{\mathfrak{#1}}
\nc{\Ac}{\Cal{A}} \nc{\Bc}{\Cal{B}} \nc{\Cc}{\Cal{C}} \nc{\Dc}{\Cal{D}} \nc{\Ec}{\Cal{E}}
\nc{\Fc}{\Cal{F}} \nc{\Gc}{\Cal{G}} \nc{\Hc}{\Cal{H}} \nc{\Ic}{\Cal{I}} \nc{\Jc}{\Cal{J}}
\nc{\Kc}{\Cal{K}} \nc{\Lc}{\Cal{L}} \nc{\Mc}{\Cal{M}} \nc{\Nc}{\Cal{N}} \nc{\Oc}{\Cal{O}}
\nc{\Pc}{\Cal{P}} \nc{\Qc}{\Cal{Q}} \nc{\Rc}{\Cal{R}} \nc{\Sc}{\Cal{S}} \nc{\Tc}{\Cal{T}}
\nc{\Uc}{\Cal{U}} \nc{\Vc}{\Cal{V}} \nc{\Wc}{\Cal{W}} \nc{\Xc}{\Cal{X}} \nc{\Yc}{\Cal{Y}}
\nc{\Zc}{\Cal{Z}}
\nc{\im}{\n{im}} \nc{\rank}{\n{rank}} \nc{\rk}{\n{rk}} \nc{\tr}{\n{tr}}
\nc{\Bb}{\bb{B}} \nc{\Cb}{\bb{C}} \nc{\Fb}{\bb{F}} \nc{\Hb}{\bb{H}} \nc{\Nb}{\bb{N}}
\nc{\Rb}{\bb{R}} \nc{\Sb}{\bb{S}} \nc{\Zb}{\bb{Z}}
\rnc{\P}{\bb{P}} \nc{\E}{\n{\bb{E}}} \nc{\Eb}{\mathop{{}\bb{E}}}
\nc{\su}{\fr{su}} \nc{\gf}{\fr{g}}  \nc{\hf}{\fr{h}} 
\nc{\RR}{\mathbf{R}} \nc{\id}{\x{id}} \nc{\SU}{\x{SU}} \nc{\Ux}{\x{U}}
\nc{\poly}{\x{poly}} \nc{\Hom}{\x{Hom}}
\nc{\ot}{\otimes} \nc{\Ot}{\bigotimes} \nc{\Oplus}{\bigoplus}
 \nc{\PT}[1]{^{\s{T}_{\!{#1}}}}
\nc{\dg}{^{\dagger}} \nc{\f}[2]{\frac{#1}{#2}}
\nc{\bra}[1]{\langle{#1}|} \nc{\ket}[1]{|{#1}\rangle}
\nc{\brak}[1]{\langle{#1}\rangle}     \nc{\ketb}[2]{|{#1}\rangle\!\langle{#2}|}
\nc{\ps}[1]{\left(#1\right)} \nc{\pss}[1]{\left[#1\right]} \nc{\psss}[1]{\left\{#1\right\}}
\nc{\xto}[1]{\xrightarrow{#1}} \nc{\dd}{\,\x{d}} \nc{\ee}{\x{e}} \nc{\ii}{\x{i}} 
\nc{\1}{\mathds{1}}  \nc{\2}{\{0,1\}}   \nc{\lims}{\varlimsup}  \nc{\limi}{\varliminf}
\nc{\ty}[1]{{\x{\tiny $#1$}}}       \nc{\too}{\!\!\to\!\!}
\nc{\mdl}{\models}              \nc{\bmdl}{=\joinrel\mathrel|}
\nc{\be}[2]{\begin{#1}#2\end{#1}}
         \nc{\Thm}[1]{\be{thm}{#1}}
     \nc{\Prop}[1]{\be{prop}{#1}}
     \nc{\Coro}[1]{\be{coro}{#1}}
        \nc{\Lem}[1]{\be{lem}{#1}}
\nc{\Eq}[1]{\be{equation}{#1}} \nc{\Eqn}[1]{\be{equation*}{#1}}
\nc{\Al}[1]{\be{align}{#1}} \nc{\Aln}[1]{\be{align*}{#1}} \nc{\Als}[1]{\be{align}{\be{split}{#1}}}
\nc{\Item}[1]{\be{itemize}{#1}} \nc{\Enum}[1]{\be{enumerate}{#1}}
\nc{\gray}[1]{{\color{gray}#1}} \nc{\red}[1]{{\color{red}#1}} \nc{\blue}[1]{{\color{blue}#1}}
\nc{\mm}[2]{\ps{\begin{array}{c}{#1}\\{#2}\end{array}}}
\nc{\mmnn}[4]{\ps{\begin{array}{cc}{#1}&{#2}\\{#3}&{#4}\end{array}}}
\nc{\ydiag}[1]{\ytableausetup{notabloids}\ydiagram{#1}}
\nc{\ytab}[1]{\ytableausetup{notabloids}\ytableaushort{#1}}

\begin{document}
\setlength{\textheight}{8.0truein}    

\runninghead{Entanglement properties of random invariant quantum states}
            {Wei Xie and Weijing Li}

\normalsize\textlineskip
\thispagestyle{empty}
\setcounter{page}{901}

\copyrightheading{22}{11\&12}{2022}{0901--0923}

\vspace*{0.88truein}

\alphfootnote

\fpage{901}

\centerline{\bf
ENTANGLEMENT PROPERTIES OF RANDOM INVARIANT QUANTUM STATES}
\vspace*{0.37truein}
\centerline{\footnotesize
WEI XIE\footnote{Corresponding author. Email: xxieww@ustc.edu.cn}}
\vspace*{0.015truein}
\centerline{\footnotesize\it School of Computer Science and Technology, University of Science and Technology of China}
\centerline{\footnotesize\it The CAS Key Laboratory of Wireless-Optical Communications, University of Science and Technology of China}
\vspace*{10pt}
\centerline{\footnotesize 
WEIJING LI}
\vspace*{0.015truein}
\centerline{\footnotesize\it School of Science, Beijing University of Posts and Telecommunications, China}
\publisher{August 25, 2021}{May 4, 2022}

\vspace*{0.21truein}

\abstracts{
Entanglement properties of random multipartite quantum states which are invariant under global $\textnormal{SU}(d)$ action are investigated.
The random states live in the tensor power of an irreducible representation of $\textnormal{SU}(d)$.
We calculate and analyze the expectation and fluctuation of the second-order R\'enyi entanglement measure of the random invariant and near-invariant states in high dimension, and reveal the phenomenon of concentration of measure the random states exhibit.
We show that with high probability a random SU($d$)-invariant state is close to being maximally entangled with respect to any bipartite cut as the dimension of individual system goes to infinity.
We also show that this generic entanglement property of random SU(2)-invariant state is robust to arbitrarily finite disturbation.
}{}{}

\vspace*{10pt}

\keywords{Quantum entanglement, random state, invariant state, concentration of measure}

\vspace*{1pt}\textlineskip    

\section{Introduction} 
Quantum entanglement as a striking and intriguing physical phenomenon exhibits fundamental nonclassical manifestation of quantum world.
Since the Einstein-Podolsky-Rosen (EPR) state was introduced in 1935~\cite{einstein1935can}, the counterintuitive feature of entanglement has been a significant theme in many fields of physics.
Nowadays as an indispensable element in quantum information and quantum physics, entangled states\textemdash especially the maximally entangled states\textemdash have found applications in quantum key distribution~\cite{bennett1984quantum,ekert1991quantum}, quantum secure direct communication~\cite{long2002theoretically}, quantum dense coding~\cite{bennett1992communication}, quantum teleportation~\cite{bennett1993teleporting}, distributed computation~\cite{fitzi2001quantum}, quantum gravity~\cite{penrose1971quantum,gyongyosi2020correlation}, etc.

As a special class of entangled states, invariant states are relevant to group representation theory.
The notion of invariant state arises generally in the study of quantum information \cite{vollbrecht2001entanglement,breuer2005entanglement}, and quantum spin systems~\cite{rovelli1995spin,baez1996spin,arnesen2001natural,breuer2005state,durkin2004resilience} as often used in the theory of loop quantum gravity~\cite{ashtekar2004background,rovelli2014covariant} and condensed matter physics.
Recall that a representation of a group $G$ (resp., a Lie algebra $\fr{g}$) on a vector space $\Hc$ is a homomorphism from $G$ (resp., $\fr{g}$) to the general linear group $\x{GL}(\Hc)$ (resp., the general linear Lie algebra $\fr{gl}(\Hc)$); either homomorphism is denoted by $\s{R}$ in this paper.
A pure state in the tensor power of an irreducible representation (irrep) of $\SU(d)$ is called {\em invariant} if it is invariant under global $\SU(d)$ action.
To be specific, for an irrep $\Hc$ of $\SU(d)$, $\ket\psi\in \Hc^{\ot n}$ is an invariant state if
\Eq{\label{def1}
\s{R}(U)^{\ot n}\ket\psi=\ket\psi \x{ for any } U\in\SU(d),
}
or equivalently,
\Eq{\label{def2}
( \s{R}(X)\ot\1^{\ot(n-1)} + 
\cdots+ \1^{\ot(n-1)}\ot\s{R}(X) )\ket\psi =0 \x{ for any } X\in\su(d).
}
For the qubit case $d=2$, the explicit form of $\s{R}(X)$ follows from \eqref{clegjlhgerjo}, and for higher dimension case the form of $\s{R}(X)$ is more complex and can be deduced from \eqref{gtbasistran}.
As a simple example, the Werner states \cite{vollbrecht2001entanglement} are invariant under all simultaneous unitary transformations $U\ot U$.
As another class of entangled states, perfect states (aka absolutely maximally entangled states~\cite{helwig2012absolute,helwig2013absolutely}) have been studied in quantum information theory, condensed matter theory and quantum gravity~\cite{hosur2016chaos,pastawski2015holographic,almheiri2015bulk}.
In particular, perfect states have close connection with optimal quantum error-correction codes \cite{raissi2018optimal}.
An $n$-partite state in $\Hc_1\ot\cdots\ot\Hc_n$ is called {\em perfect} if with respect to any bipartition of the $n$ systems it is a maximally entangled state of size equal to the smaller dimension of this bipartition~\cite{pastawski2015holographic}.
It can be seen that the Greenberger-Horne-Zeilinger state $\f{1}{\sqrt2}(\ket{000}+\ket{111})$ is perfect while the $W$ state $\f{1}{\sqrt3}(\ket{100}+\ket{010}+\ket{001})$ is not.

The relation between the two classes of states (i.e., invariant states and perfect states) is not entirely clear, despite much relevant research for various settings.
The entanglement properties of bipartite invariant states with respect to SU(2) and SO(3) were analyzed using various entanglement criteria~\cite{schliemann2003entanglement,schliemann2005entanglement,breuer2005entanglement}.
The existence and construction of invariant perfect states were studied in \cite{li2017invariant}.
The SU(2)-invariant state which is chosen uniformly at random  was proved to be asymptotically perfect as the representation dimension goes to infinity \cite{li2018random}.
The asymptotic perfectness is quite similar to the well-known phenomenon that a high-dimensional random bipartite pure state is nearly maximally entangled~\cite{hayden2006aspects}.
Various concentration of measure phenomena in high dimension have attracted great attention \cite{hayden2004randomizing,hayden2006aspects,collins2016random} and inspired applications in the field of quantum information and quantum physics such as the proof of nonadditivity of Holevo capacity \cite{dahlsten2014entanglement,hastings2009superadditivity}.

In the present work we further investigate the relation between invariant states and perfect states in the asymptotic setting.
An interesting question is whether and, if so, to what extent, the property of asymptotic perfectness of a random invariant state is robust to finite disturbation. 
To be specific, if a random state may be slightly disturbed (but not kept completely invariant) under global $\SU(2)$ action, is it still asymptotically perfect?
We call this class of states {\em near-invariant}, which generalizes previous notion of invariant states.
We show that a random high-dimensional near-invariant state is typically perfect, which means that the asymptotic perfectness of random SU(2)-invariant states is robust to {\em arbitrarily} finite disturbance.

We also generalize the study of $\SU(2)$-invariance to $\SU(d)$-invariance by showing that the typicality of entanglement holds for this new symmetry.
In physics the group SU($d$) for $d\ge3$ can emerge as an effective model for describing systems with relevant symmetries
\cite{gauthe2019tensor}.
In quantum information science, random states and circuits with respect to SU($d$) have a great many applications.
If all that we know about a quantum system is the representation space it lives in, it is reasonable to regard it as a uniformly random model.
In this case the expectation of the random states is an invariant state and one can use this invariant state to estimate certain properties such as entanglement measure of the system \cite{watrous2018theory}.

The structure of this paper is as follows.
In Section \ref{sec3} we show that the asymptotic perfectness of a random SU(2)-invariant state is robust to arbitrarily finite disturbance by analyzing its 2-R\'enyi entanglement entropy.
In Section \ref{sec4} we investigate the entanglement in random SU($d$)-invariant state and show that in large dimension it is perfect with high probability.
For convenience of the reader, in Appendix \hyperref[AppA]{A} we briefly review the representation theory of $\su(2)$ and $\su(d)$ with some notation fixed.
For readability, the lengthy proofs of several results in Sections \ref{sec3} and \ref{sec4} are postponed into Appendices \hyperref[AppB]{B} and \hyperref[AppC]{C}, respectively.

\section{Random near-invariant states}\label{sec3}
In this Section we investigate the robustness of the asymptotic perfectness of random states which are invariant under SU(2) action by analyzing the entanglement measure of near-invariant states.
Roughly speaking, an $n$-partite state $\ket\psi$ is called {\em near-invariant} with respect to SU(2) if 
\Eq{
\s{R}(U)^{\ot n}\ket\psi\approx\ket\psi \x{ for any } U\in\SU(2).
}

To analyze its entanglement, we need some knowledge of representation theory of unitary group,
which is briefly introduced in Appendix \hyperref[AppA]{A} (see \cite{fulton2013representation,hall2015lie,goodman2009symmetry} for a systematic exposition).
There is a one-to-one correspondence between representations of unitary group and its Lie algebra, and we focus on the latter for simplicity.
The following two formulas are basic results from representation theory \cite{weyl1939classical} and will be used later:
\Eq{\label{twirl1}
\int U^{\ot 2} X(U\dg)^{\ot 2}\dd U=c_1\1+c_2W
}
where $\x{d}U$ denotes the Haar measure on $\SU(d)$, $W$ is the swap operator, and
\Eq{\label{twirl2}
\int\!U^{\ot n} \ketb{\psi}{\psi}(U\dg)^{\ot n}\dd U\!=\!\f{1}{d(d+1)\cdots(d\!+\!n\!-\!1)}\!\sum_{\pi\in S_n}\!\!W_\pi
}
where $W_\pi$ is the operator permutating $n$ tensor factors according to $\pi\in S_n$.

We denote by $V_{(s)}$ the irrep of $\SU(2)$ of dimension $2s+1$ for any nonnegative integer or half-integer $s$.
We now characterize near-invariant states in $V_{(s)}^{\ot n}$.
The space $V_{(s)}^{\ot n}$, as a new representation of $\su(2)$, has the following decomposition based on Clebsch-Gordan transform
\Eq{\label{sec33-trans}
  V_{(s)}^{\ot n} = \Oplus_{j\le ns} V_{(j)}^{\oplus N(n,j)} \,,
}
where $N(n,j)$ is the multiplicity of irrep $V_{(j)}$. 
A basis of the right-hand side of Eq.\ \eqref{sec33-trans} consists of $\ket{j,m,\gm}$ with $j\in\{0,1,\dots,ns\}$ (when $s$ is a half-integer, $j\in\{\f{1}{2},\f{3}{2},\dots,ns\}$), $\gm\in\{1,2,\dots,N(n,j)\}$ and $m\in\{j,j-1,\dots,-j\}$, which is called the {\em coupled} basis.
As a special case, by Eq.\ \eqref{def1}, a state $\ket\psi$ is invariant if and only if it is a superposition of states $\ket{j=0,m=0,\gm}$.
The invariant perfect state exists for $n=2$ or $n=3$ but not for $n=4$ \cite{li2017invariant}.
Notice that SU(2)-invariant state does not exist in $n$ systems of $s$-spin when $s$ is a half-integer and $n$ is odd, and the introduction of near-invariant state can address this issue.

Considering the decomposition~\eqref{sec33-trans}, we define the {\em near-invariant state space} $\Hc_{s,n,j_0}^\x{inv}$ as the unit sphere of 
\Eq{\label{def-near-inv}
\x{span}\{\ket{j,m,\gm}:j\le j_0,-j\le m\le j,1\le\gm\le N(n,j)\}
}
where $j_0\ge0$ is arbitrarily finite.
We now analyze the perfectness of a uniformly random state in $\Hc_{s,n,j_0}^\x{inv}$.
Consider the bipartition of the $n$ systems such that Alice holds $p$ systems and Bob holds the rest $q:=n-p$ systems with $p\le q$.
Thus
\Eq{\label{ch3:shkaf}
V_{(s)}^{\ot n}=\bigg( \Oplus_{j_1\le ps} V_{(j_1)}^{\oplus N(p,j_1)} \bigg)
            \Ot \bigg( \Oplus_{j_2\le qs} V_{(j_2)}^{\oplus N(q,j_2)} \bigg) \,.
}

The space $V_{(s)}^{\ot n}$ on the left-hand side of~\eqref{ch3:shkaf} has a basis $\{ \ket{j,m,\gm}\}$, while the space on the right-hand side has a basis $\{ \ket{j_1,m_1,\alp_{j_1}} \ot \ket{j_2,m_2,\bt_{j_2}} \}$ where $1\le \alp_{j_1}\le N(p,j_1)$ and $1\le \bt_{j_2}\le N(q,j_2)$.
It follows from \eqref{def-near-inv} that the near-invariant state space $\Hc_{s,n,j_0}^\x{inv}$ has a basis $\{ \ket{j m j_1 j_2 \alp_{j_1} \bt_{j_2}} : j\le j_0,-j\le m\le j,j_1\le ps,j_2\le qs,1\le \alp_{j_1}\le N(p,j_1),1\le \bt_{j_2}\le N(q,j_2) \}$, where
\Eq{\label{cgc3}
  \ket{j m j_1 j_2 \alp_{j_1} \bt_{j_2}} = \sum \ket{j_1m_1\alp_{j_1}}_\ty{A} \ket{j_2m_2\bt_{j_2}}_\ty{B} C^{j_1,j_2,j}_{m_1,m_2,m}
}
with the sum over $m_1,m_2$ satisfying $m_1+m_2=m$ and $C$ denoting the Clebsch-Gordan coefficients.
Eqs.\ \eqref{ch3:shkaf} and \eqref{cgc3} are useful for our analysis of multipartite quantum entanglement.
In fact, similar decompositions of representations have been playing an important role in the study of entanglement, such as the classification of entanglement \cite{gour2013classification} and concentration of entanglement \cite{matsumoto2007universal}.

Let $\ket\phv_\ty{AB}$ be a uniformly random state in $\Hc_{s,n,j_0}^\x{inv}$, and $\phv_\ty{A}$ be its reduced state on system $A$.
By definition, the 2-R\'enyi entanglement entropy of $\ket\phv_\ty{AB}$
is
\Eq{
H_2(\phv_\ty{AB}):=-\ln\tr(\phv_\ty{A}^2).
}
Then 
\Als{\label{fvdghgh}
  H_2(\phv_\ty{AB}) &= -\ln\tr( W_\ty{\!AA'} (\phv_\ty{A}\ot\phv_\ty{A'}) ) \\
  &= -\ln\tr( (W_\ty{\!AA'}\ot\1_\ty{BB'}) (\phv_\ty{AB}\ot \phv_\ty{A'B'}) ) \,,
}
where $\phv_\ty{AB}$ and $\phv_\ty{A'B'}$ are the same state on isomorphic systems $AB$ and $A'B'$, respectively.
Notice that the 2-R\'enyi entanglement entropy of a maximally entangled state on $AB$ is $H_\x{max}:=\ln(2s+1)^{p}$.
We are concerned with how close the random invariant state $\phv_\ty{AB}$ is to a maximally entangled state.
The closeness is measured by the ratio
\Eq{\label{ch3:closeness}
  \eta(\phv_\ty{AB}):=\f{ H_2(\phv_\ty{A}) }{ H_\x{max} }  \,.
  }
If the ratio for $\phv_\ty{AB}$ is close to 1, then obviously the state is close to being maximally entangled.

As it is difficult to calculate the average of $H_2(\phv_\ty{A})$ directly, we will estimate the average of $\ee^{-H_2(\phv_\ty{A})}$.
It follows from Eq.\ \eqref{fvdghgh} that
\Eqn{
\E_\phv \ee^{-H_2(\phv_\ty{A})} = \E_\phv \tr(\phv_\ty{A}^2) = \tr( W_\ty{AA'} \E_\phv(\phv_\ty{AB}\ot \phv_\ty{A'B'}) ) \,.
}
Using Werner twirling formula \eqref{twirl1}, we have 
\Eq{
  \E_\phv (\phv_\ty{AB}\ot \phv_\ty{A'B'}) = \f{ \1_\x{inv}^{\ot2}+W_\x{inv} }{ d_\x{inv}^2+d_{\x{inv}} } \,,
  }
where $\1_\x{inv}$ is the projector onto the near-invariant space $\Hc_{s,n,j_0}^\x{inv}$, $d_\x{inv}=\tr(\1_\x{inv})$ is its dimension, and $W_\x{inv}$ is the swap operator on $(\Hc_{s,n,j_0}^\x{inv})^{\ot 2}$.

It turns out that the random variable $H_2(\phv_\ty{A})$ is highly concentrated around $-\ln \E \ee^{-H_2(\phv_\ty{A})}\sim p\ln s$ with small fluctuation for large dimension:

\Thm{\label{ch3-thm}
  A random near-invariant state with respect to $\SU(2)$ is asymptotically perfect as the dimendion of individual system goes to infinity.
  
  To be specific, for any $p,q$ such that $q\ge p\ge 2$ and $q\ge 3$ and any $\dt>0$, a random near-invariant state $\phv_\ty{AB} \in \Hc_{s,n,j_0}^\x{inv} \subseteq V_{(s)}^{\ot p}\ot V_{(s)}^{\ot q}$ satisfies $\Pr(|\eta(\phv_\ty{AB})-1|\ge \dt) \to 0$ as $s\to\infty$.
}

The proof of Theorem \ref{ch3-thm} requires the following two Propositions.
In the sequel we write $f\lesssim g$ for two positive functions $f$ and $g$ of $s$ if $f(s)\le cg(s)$ holds for some constant $c$ and any sufficiently large $s$.

\Prop{\label{ch2-prop-expec}
  Let $q\ge p\ge 2$.
  For random near-invariant state $\phv_\ty{AB}$, as $s\to\infty$,
  $$
  s^{-p}\lesssim \E_\phv \ee^{-H_2(\phv_\ty{A})}\lesssim s^{-p}\ln s \,.
  $$
}

\Prop{\label{ch2-prop-higher}
  Let $q\ge p\ge 2$ and $q\ge 3$.
  For random near-invariant state $\phv_\ty{AB}$, as $s\to\infty$,
  $$
  \f{ \E_\phv (\tr\phv^2_\ty{A})^2 }{ (\E_\phv\tr\phv_\ty{A}^2)^2 }-1 \lesssim s^{-1} \,.
  $$
}

Proposition \ref{ch2-prop-expec} gives upper and lower bounds on the average of the second moment of random state $\phv_\ty{A}$.
Proposition \ref{ch2-prop-higher} bounds the fluctuation of this second moment.
The proofs of the two Propositions are given in Appendix \hyperref[AppB]{B}.
By invoking the two Propositions, Theorem~\ref{ch3-thm} is proved as follows.

\proof{of Theorem~\ref{ch3-thm}.
Denote $K:=-\ln\E\ee^{-H_2(\phv_\ty{A})}$.
Then by Proposition~\ref{ch2-prop-expec}, there exist constants $c,C$ such that $p\ln s-\ln\ln s-\ln C \le K\le p\ln s-\ln c$.
Thus $\f{K}{H_\x{max}}\to 1$ as $s\to\infty$.
So for any $\dt>0$ and $s$ large enough,
\Eq{\label{falhse}
  \Big|\f{K}{H_\x{max}}-1\Big|\le\f{\dt}{2} \,.
}

We have
\Als{\label{vnsangobwrbgr}
  \Pr & \Big(\Big|\f{H_2(\phv_\ty{A})}{H_\x{max}}-1\Big|\ge \dt\Big) \\
  &= \Pr\Big(\Big|\f{H_2(\phv_\ty{A})-K}{H_\x{max}}+\f{K}{H_\x{max}}-1\Big|\ge\dt\Big) \\
  & \le \Pr(|H_2(\phv_\ty{A})-K|\ge H_\x{max}\dt/2) \\
  & \le \Pr\Big(\Big|\f{\tr\phv_\ty{A}^2}{\E\tr\phv_\ty{A}^2}-1\Big|\ge\epsl\Big) \\
  & \le \f{1}{\epsl^2} \Big(\f{ \E (\tr\phv^2_\ty{A})^2 }{ (\E \tr\phv_\ty{A}^2)^2 }-1\Big) \\
  & \lesssim \f{1}{s\epsl^2}\to 0 \,,
}
where $\epsl:=1-\ee^{-H_\x{max}\dt/2}$, the second line uses Eq.\ \eqref{falhse}, the third line uses the fact that if $|\ln x|\ge t$ then $|x-1|\ge1-\ee^{-t}$, the fourth line uses Markov inequality, and the last line is due to Proposition~\ref{ch2-prop-higher}.
}

Indeed, it can be seen from this proof that $\dt$ can be chosen arbitrarily small as long as $\dt\gtrsim s^{-1/2-\epsl}$ holds for any positive $\epsl$.

The generic entanglement property of random near-invariant states is now established by Theorem \ref{ch3-thm}, showing that this property is rubust to arbitrarily finite perturbance since this result is {\em independent} of the value of $j_0$.
Compared with previous proof ideas in~\cite{hayden2016holographic,li2018random}, our proof is simplified as the uncoupled basis is not used in the calculation.

\section{Invariant states of higher degree}\label{sec4}

In this Section we study the typical entanglement in invariant states with respect to $\SU(d)$, generalizing previous results on $\SU(2)$.
Although the basic idea of representation of $\su(d)$ for $d\ge 3$ is similar to that of $\su(2)$, substantive difference exists as the former involves complex structure of roots and weights~\cite{hall2015lie}.
A brief introduction to $\su(d)$ representation theory and relevant notions are given in Appendix \hyperref[AppA]{A}.

As a simple example, invariant state exists in $(\Cb^d)^{\ot n}$ when $n$ is divisible by $d$.
Indeed, consider the Schur-Weyl decomposition $(\Cb^d)^{\ot n}=\Oplus_{\ld\in\x{Par}(n,d)}V_\ld\ot \Kc_\ld$, where $V_\ld$ is the Weyl module of $\su(d)$ and $\Kc_\ld$ is the Specht module of $S_n$.
Here and in what follows, we use $\x{Par}(n)$ to denote the set of all partitions of $n$, and write $\ld\vdash n$ for $\ld\in\x{Par}(n)$.
If $n$ is divisible by $d$, the invariant state space in $(\Cb^d)^{\ot n}$ always exists uniquely by definition~\eqref{def2}, which is $V_\ld\ot\Kc_\ld$ for $\ld$ being $(k,k,\dots,k)$ with $k=n/d$.

We now investigate the entanglement property of random SU($d$)-invariant $n$-partite state and show its asymptotic perfectness similarly to the case of SU(2)-invariance.
We denote by $Y_{(s)}$ the symmetric subspace of $(\Cb^d)^{\ot s}$ for fixed $d$, which is an irrep of $\su(d)$.
In the same way as the SU(2)-invariance case, we consider the bipartition that Alice and Bob hold $p$ and $q$ systems respectively where $p\le q$.
We again use the following ratio to measure the closeness of a given state $\phv_\ty{AB}$ to maximally entangled state:
\Eq{
\eta(\phv_\ty{AB}):=\f{ H_2(\phv_\ty{A}) }{ H_\x{max} } \,,
} 
where the 2-R\'enyi entropy of the maximally entangled state is
\Eq{\label{vnjksdnvjilv}
H_\x{max}:=\ln\big(\dim Y_{(s)}^{\ot p}\big) = \ln\binom{s+d-1}{d-1}^p \simeq \ln s^{(d-1)p} \,.
}

The main result of this Section is as follows.

\Thm{\label{fwlngearnba}
  For $p\ge d\ge 2$, the random invariant state in $Y_{(s)}^{\ot n}$ for $\SU(d)$ is asymptotically perfect as the dimension of individual system goes to infinity.
  
  To be specific, for any $p\le q$ and any $\dt>0$, a random invariant state $\phv_\ty{AB} \in \Hc_{s,n}^\x{inv} \subseteq Y_{(s)}^{\ot p}\ot Y_{(s)}^{\ot q}$ satisfies $\Pr(|\eta(\phv_\ty{AB})-1|\ge \dt) \to 0$ as $s\to\infty$.
}

Before turning to its proof we study the condition for existence of invariant state in $V_\ld\ot V_\mu$ which is relevant to dual partitions and dual Gelfand-Tsetlin (GT) patterns (see Appendix \hyperref[AppA]{A} for details of GT patterns).
We say two partitions $\ld:=(\ld_1,\dots,\ld_d)$ and $\mu:=(\mu_1,\dots,\mu_d)$ are dual to each other if $\ld_k+\mu_{d+1-k}=\ld_1+\mu_{d}$ for each $k\in[d]:=\{1,2,\dots,d\}$.
Similarly, we say two GT patterns $\bm\ld:=(\ld_k^l)_{l\in[d],k\in[l]}$ and $\bm\mu:=(\mu_k^l)_{l\in[d],k\in[l]}$ are dual to each other if $\ld_k^l+\mu_{l+1-k}^l=\ld_1^l+\mu_l^l$ for each $l\in[d]$ and $k\in[l]$.
Given $\ld=(\ld_1,\dots,\ld_d)\in\x{Par}(ps,d)$, the dual of $\ld$ in $\x{Par}(qs,d)$ is unique and denoted by $\ld_*$, and moreover
the dual of $\bm\ld\in\x{GT}(\ld)$ in $\x{GT}(\mu)$ with $\mu\in\x{Par}(qs,d)$ is also unique and denoted by $\bm{\ld_*}$.
The condition for the existence of invariant state in tensor product of two irreps is specified in the following Lemma of which the proof is postponed into Appendix \hyperref[AppC]{C}.

\Lem{\label{cnjksnvkvbad}
  Let $\ld:=(\ld_1,\dots,\ld_d)$ and $\mu:=(\mu_1,\dots,\mu_d)$ be two partitions.
  As representation of $\su(d)$, $V_\ld\ot V_\mu$ contains invariant state space if and only if $\ld$ and $\mu$ are dual to each other.
  Further, for $\bm\ld\in\x{GT}(\ld)$ and $\bm\mu\in\x{GT}(\mu)$, $C_{\bm\ld,\bm\mu}^{0}\ne 0$ if and only if $\bm\ld$ and $\bm\mu$ are dual to each other.
}

We now turn to the proof of Theorem \ref{fwlngearnba}.
For any integer $k$ and $\ld\in\x{Par}(ks)$, the multiplicity of $V_\ld$ in decomposition of $Y_{(s)}^{\ot k}$ (the multiplicity of $\ld$ in $(s)^{\ot k}$ for short) is denoted by $N(\ld)$.
Consider the decomposition
\Aln{
Y_{(s)}^{\ot n} &= Y_{(s)}^{\ot p} \ot Y_{(s)}^{\ot q} \\
&=\Big(\Oplus\nolimits_{\ld\vdash ps} V_\ld^{\oplus N(\ld)}\Big) \Ot \Big(\Oplus\nolimits_{\mu\vdash qs} V_{\mu}^{\oplus N(\mu)} \Big) \,.
}
It suffices to focus on the dual pairs of partitions and patterns in $Y_{(s)}^{\ot p}\ot Y_{(s)}^{\ot q}$.
In the rest of this Section we assume that $\f{p+q}{d}s$ is an integer so that $Y_{(s)}^{\ot p}\ot Y_{(s)}^{\ot q}$ contains invariant states.
For any dual pair of partitions $\ld\in\x{Par}(ps,d)$ and $\mu\in\x{Par}(qs,d)$ and for $1\le \alp_\ld \le N(\ld)$ and $1\le \bt_{\mu}\le N(\mu)$,
\Eq{\label{decomp-cgc-sud}
  \ket{\ld,\alp_\ld,\bt_\mu} = \sum C_{\bm\ld,\bm\mu}^0 \ket{\bm\ld,\alp_\ld}_\ty{A} \ket{\bm\mu,\bt_\mu}_\ty{B} \,,
}
where the sum is over dual pairs of patterns $\bm\ld\in\x{GT}(\ld)$ and $\bm\mu\in\x{GT}(\mu)$, is an invariant state.

Using the proof idea as used in Theorem~\ref{ch3-thm} we need to estimate the expectation $\E\ee^{-H_2(\phv_\ty{A})}=\E\tr\phv_\ty{A}^2$ and give a bound on its fluctuation $\f{\E(\tr\phv_\ty{A}^2)^2}{(\E\tr\phv_\ty{A}^2)^2}$ using Markov inequality.
We have 
\Eq{\label{cnwoehfveoe}
\E\tr\phv_\ty{A}^2 = \f{\tr(W_\ty{AA'}(\1_\x{inv}^{\ot 2}+W_\x{inv}))}{d_\x{inv}^2+d_\x{inv}} \,,
}
where
\Eq{
d_\x{inv}=\sum_{\ld\vdash ps,\ld_*\vdash qs} N(\ld)N(\ld_*)
}
is the dimension of invariant state space.

The states of form~\eqref{decomp-cgc-sud} constitute an orthonormal basis of the invariant state space, so $\1_\x{inv}^{\ot 2}$ is a sum of $\big( \ket{\ld,\alp_\ld,\bt_\mu } \ket{ \ld',\alp'_{\ld'},\bt'_{\mu'} } \big)
\big( \bra{ \ld,\alp_\ld,\bt_\mu} \bra{ \ld',\alp'_{\ld'},\bt'_{\mu'} } \big)$
over $\ld,\alp_\ld,\bt_\mu,\ld',\alp'_{\ld'},\bt'_{\mu'}$ such that $\ld\vdash ps$ is dual to $\mu\vdash qs$.
Inserting~\eqref{decomp-cgc-sud} into the expression of $\1_\x{inv}^{\ot 2}$, we have
\Aln{
\tr (W_\ty{AA'}\1_\x{inv}^{\ot 2}) &= \sum_{ \ld,\alp_\ld,\bt_{\ld_*},\bt'_{\ld_*} } \sum_{\bm\ld} (C_{\bm\ld,\bm{\ld_*}}^0)^4 \\
&= \sum_\ld  N(\ld)N(\ld_*)^2 \f{1}{\dim V_\ld} \,,
}
since $(C_{\bm\ld,\bm{\ld_*}}^0)^2=(\dim V_\ld)^{-1}$.

Similarly,
\Eq{
  \tr (W_\ty{AA'}W_\x{inv}) = \sum_\ld  N(\ld)^2N(\ld_*) \f{1}{\dim V_\ld} \,.
}

An estimate of the multiplicities $N(\ld),N(\ld_*)$ of irreps in decomposition of $Y_{(s)}^{\ot p}\ot Y_{(s)}^{\ot q}$ with $p\le q$ is given in Lemmas \ref{Lemma10} and \ref{Lemma11} in Appendix \hyperref[AppC]{C}.

For large $n$ and constant $k$, we have $|\x{Par}(n,k)|\simeq n^{k-1}$ since $\f{1}{k!}|\x{Type}(n,k)|\le\x{Par}(n,k)\le|\x{Type}(n,k)|$ and $|\x{Type}(n,k)|=\binom{n+k-1}{k-1}$.
Here, $\x{Type}(n,k)$ denotes the set of all $k$-tuples of nonnegative integers that sum to $n$.
Using Weyl dimension formula, $\dim V_\ld\simeq s^{h(h-1)/2}$ for $h:=\min\{p,d\}$.
Taking together these estimates, we have 
\[
\E\tr\phv_\ty{A}^2 \simeq s^{-p(d-1)} \,.
\]
Then by invoking \eqref{vnjksdnvjilv}, $\f{-\ln\E\ee^{-H_2(\phv_\ty{A})}}{H_\x{max}}\to1$ as $s\to\infty$.
A lengthy calculation shows
\[
\f{\E(\tr\phv_\ty{A}^2)^2}{(\E\tr\phv_\ty{A}^2)^2}-1 
\lesssim  s^{(d-1)(-p-q+d+2)} \,.
\]
Under the condition $q\ge p\ge d\ge 3$, this upper bound vanishes.
Consequently, using \eqref{vnsangobwrbgr}, the random invariant state is perfect with high probability.
The technical derivation can be found in Lemma \ref{josdhbdfhbkvkjsadn} in Appendix \hyperref[AppC]{C}.
Now we have finished the proof of Theorem \ref{fwlngearnba}.

By similar calculation, when $p<d$, we have that $\E\tr\phv_\ty{A}^2 \gtrsim s^{-p(p-1)}$, and thus $\f{-\ln\E\tr\phv_\ty{A}^2}{H_\x{max}} \le \f{p-1}{d-1} < 1$ for large $s$.
Thus the preceding derivation is invalid for the case where the numbers $p,q$ of systems Alice and Bob hold are relatively small compared with the local dimension $d$, in that $\tr\phv_\ty{A}^2$ is so small that $-\ln\E \tr\phv_\ty{A}^2$ is far from $-\E \ln\tr\phv_\ty{A}^2$, as the second-order derivative of $-\ln x$ is large for small positive $x$.

\section{Conclusion and discussion}\label{sec5}
We have rigorously studied the typical entanglement in random multipartite states which are kept invariant or finitely disturbed under product representation of special unitary group, extending and generalizing previous results on similar states.
To be specific, we have calculated and analyzed the expectation and fluctuation of 2-R\'enyi entanglement entropy of random SU($d$)-invariant states in the large dimension limit.
We have shown in a precise way that these random invariant states are perfect, i.e., highly entangled with respect to any bipartite cut, and that the perfectness of random SU(2)-invariant state is robust to arbitrarily finite disturbance.

Our study subsumes existing results
such as in \cite{li2017invariant,li2018random} in the following two respects:
(1) When $j_0$ is set to be 0, the near-invariant state becomes SU(2)-invariant;
(2) In our discussion of random high-degree invariant states, when $d$ is set to be 2, $Y_{(s)}$ becomes an irrep of SU(2) and the random state becomes SU(2)-invariant.
In this paper we consider arbitrary values of $j_0$ and $d$, and show that the asymptotic perfectness holds for the extended random state models.
In addition, compared with previous works, our derivation simplifies the calculation by not involving the uncoupled basis while asymptotic estimation of relevant variables is given accurately.

The model of high-dimensional random state has been playing a significant
role in the theory of quantum information and quantum computing (see e.g., \cite{harrow2004superdense,hayden2006aspects,bremner2009random,hastings2009superadditivity,hayden2016holographic}).
Many of these applications are based on a fundamental fact that random quantum states are typically extremely highly entangled \cite{hayden2004randomizing,hayden2006aspects}.
The breakthrough proof \cite{hastings2009superadditivity} of nonadditivity of Holevo capacity of a quantum channel pair is a remarkable example, showing that entanglement can increase the capacity of quantum channels to send classical information.
The bipartite maximally entangled state is used in the proof as it is invariant under $U\ot U^*$ for any unitary $U$.
It is not known, however, whether tripartite and multipartite entangled states can be used to increase Holevo capacity.
The random invariant state studied in this work, as an
extended model of ordinary random state, has potential application to the study of multipartite nonadditivity of Helevo capacity, since the tripartite state which is invariant under $U_1\ot U_2\ot U_3$ for $U_i$ being some representations of unitary group is an appealing choice of the input state of random channels.

As another research direction, one can study the entanglement of mixed invariant states \cite{schliemann2003entanglement}, and the random tensor networks \cite{hayden2016holographic,qi2017holographic} with local SU($d$) symmetry.
We believe that the phenomenon of concentration of measure as revealed in our work will pave the way toward
further research on application of entanglement in quantum information science as the study of ordinary random state did.

\nonumsection{Acknowledgements}
\noindent
We thank an anonymous reviewer for helpful suggestions.
This work was supported by the Fundamental Research Funds for the Central Universities (Grant No.\ WK2150110023), Anhui Initiative in Quantum Information Technologies (Grant No.\ AHY150100), National Natural Science Foundation of China (Grant No.\ 62102388 and 62002333), and Innovation Program for Quantum Science and Technology (Grant No.\ 2021ZD0302902).

\nonumsection{References}
\noindent

\bibliographystyle{ieeetr}

\appendix{ Basics of representation theory}\label{AppA}

For convenience of the reader, we review in this Appendix some basics of the representation theory of $\su(2)$ and $\su(d)$, which can be found in textbooks such as \cite{fulton2013representation,hall2015lie,goodman2009symmetry}.
Since the Lie algebra $\fr{sl}(d)$ is the complexification of $\su(d)$, there is a one-to-one correspondence of their irreps.
It is convenient for our purpose to work with $\fr{sl}(d)$ or $\fr{gl}(d)$.

Let $J_x,J_y,J_z$ be the angular momentum operators such that $[J_k,J_l]=\ii \epsilon_{klm} J_m$ where $k,l,m\in\{x,y,z\}$ and $\epsilon_{klm}$ is the Levi-Civita symbol ($\hbar=1$ is taken).
Denote $J_\pm:=J_x\pm \ii J_y$.
For any positive integer $k$, there exists a $k$-dimensional irrep of $\su(2)$.
Let $j:=(k-1)/2$ and let $V_{(j)}:=\x{span}\{\ket{j,j},\ket{j,j-1},\dots,\ket{j,-j}\}$ denote this $k$-dimensional irrep.
Any $k$-dimensional irrep of $\su(2)$ is isomorphic to $V_{(j)}$.
The action of $\su(2)$ on $V_{(j)}$ can be written as
\Als{\label{clegjlhgerjo}
J_z\ket{j,m} &= m \ket{j,m}  \\
J_\pm \ket{j,m} &= \sqrt{(j\mp m)(j\pm m+1)} \ket{j,m\pm 1} \,.
}

Recall that the tensor product of irreps of $\su(2)$ is decomposed into a direct sum of irreps:
\Eq{
  V_{(j_1)}\ot V_{(j_2)} = \Oplus_{|j_1-j_2|\le j\le j_1+j_2} V_{(j)} \,.
}
The basis of $\Oplus_j V_{(j)}$ can be written as
\Eq{\label{ch3:expansion}
\ket{(j_1j_2)jm} =\sum_{m_1,m_2}\ket{j_1m_1j_2m_2} \brak{j_1m_1j_2m_2 | (j_1j_2)jm} \,,
}
where $\brak{j_1m_1j_2m_2 | (j_1j_2)jm}=:C_{m_1,m_2,m}^{j_1,j_2,j}$ are {\em Clebsch-Gordan coefficients} (CGCs).

A closed-form expression for the CGCs of $\su(2)$, known as Racah's formula~\cite{racah1942theory}, is 
\Eq{\label{racah}
C^{j_1,j_2,j}_{m_1,m_2,m}= \dt_{m,m_1+m_2} \sqrt{B_1B_2} \sum\nolimits_k (-1)^k f_k^{-1} \,, 
}
where $B_1:=(2j+1)(j+j_1-j_2)!(j-j_1+j_2)!(j_1+j_2-j)!/(j_1+j_2+j+1)!$, $B_2:=(j+m)!(j-m)!(j_1+m_1)!(j_1-m_1)!(j_2+m_2)!(j_2-m_2)!$, and $f_k:=k!(j_1+j_2-j-k)!(j_1-m_1-k)!(j_2+m_2-k)!(j-j_2+m_1+k)!(j-j_1-m_2+k)!$.

We now introduce the representation theory for $\fr{gl}(d)$ for convenience of the reader.
An irrep of $\fr{gl}(d)$ is labeled by its highest weight, i.e., a sequence $\ld$ of $d$ nonincreasing integers.
The irrep of highest weight $\ld=(\ld_1^d,\dots,\ld_d^d)$ is denoted by $V_\ld$.
As irreps of $\fr{sl}(d)$ and $\su(d)$, $V_{(\ld_1^d,\dots,\ld_d^d)}$ and $V_{(\ld_1^d+c,\dots,\ld_d^d+c)}$ are equivalent for any integer $c$.
So irreps of $\fr{sl}(d)$ (and $\su(d)$) can be labeled by sequences of $d$ nonincreasing integers with the last integer being zero.

The basis states of the irrep $V_\ld$ can be labeled by the {\em Gelfand-Tsetlin (GT) patterns}, which are arrays of integers, of the following form
$$
\bm{\ld}=\left(\begin{array}{l}
\ld_1^d \phantom{wwv} \ld_2^d \phantom{ww} \cdots \phantom{ww} \ld_d^d \\
\phantom{vv} \ld_1^{d-1} \phantom{vv} \cdots \phantom{vv} \ld_{d-1}^{d-1} \\
\phantom{www} \ddots \phantom{wwv} \iddots \\
\phantom{wwwwwv} \ld_1^1
\end{array}\right)
$$
satisfying the interlacing condition
\Eqn{
  \ld^l_1 \ge \ld^{l-1}_1 \ge \ld^l_2 \ge \ld^{l-1}_2 \ge \cdots\ge \ld^{l-1}_{l-1} \ge \ld^l_l 
}
for each $l\in\{2,\dots,d\}$.

Let $\x{GT}(\ld)$ denote the set of GT patterns with the top row being $\ld$.
Each GT pattern $\bm\ld$ corresponds to a basis state, denoted $\ket{\bm\ld}$, and via the branching rule these states constitute an orthonormal basis of $V_\ld$, i.e., $V_\ld=\x{span}\{\ket{\bm\ld}: \bm\ld\in\x{GT}(\ld)\}$.

The $k$-th element in row $l$ of $\bm\ld$ is denoted by $\bm\ld^l_k$ or $\ld^l_k$.
Let $\ld^l$ or $\bm\ld^l$, $l=1,\dots,d$, denote the $l$-th row $(\ld_1^l,\dots,\ld_l^l)$;
the top row is written as $\ld^d$ or $\ld$.
Let $\x{GT}(\ld;\ld^{d-1})$ denote the set of GT patterns with the top two rows being $\ld$ and $\ld^{d-1}$.
We use the bold letters $\bm\ld,\bm\mu,\bm\nu$ to denote GT patterns with the top row being $\ld,\mu,\nu$, respectively.

In the same spirit as that a partition can be represented by a Young diagram, a GT pattern $\bm\ld$ can be represented by a semistandard Young tableau of shape $\ld$ and alphabet $\{1,\dots,d\}$, e.g.,
\begin{figure}[H] \centering \begin{tikzpicture}
  \node[right,xshift=0cm,yshift=0cm] at (0,0)
  {$\left(\begin{array}{ccccc}
    5 & & 3 & & 2 \\
  & 4 & & 2 & \\
  &   & 3 &   & 
  \end{array}\right)$
  };
  \node[right,xshift=3.3cm,yshift=0cm] at (0.4,0) {$\iff$};
  \node[right,xshift=5cm,yshift=0cm] at (0,0) {\scalebox{1.2}{\ytab{11123,223,33}}};
  \node[right,xshift=7.2cm,yshift=-.2cm] at (0,0) {.};
\end{tikzpicture} \end{figure}

The action of $\fr{gl}(d)$ on $V_\ld$ is introduced as follows.
Let $\bm\ld+ 1^{K,L}$ denote the array of integers such that $\bm\ld+1^{K,L}$ and $\bm\ld$ have the same elements except that the $K$-th element in row $L$ of $\bm\ld+1^{K,L}$ is $\ld_K^L+ 1$.
The array $\bm\ld- 1^{K,L}$ is defined in the same fashion.
$\bm\ld\pm 1^{K,L}$ may not be a valid GT pattern.
Denote by $E^{i,j}$ the matrix with a one in the $j$-th entry in row $i$ and with zero elsewhere.
For $1\le l\le d-1$, a basis state $\ket{\bm\ld}$ is acted by $\fr{gl}(d)$ as
\Als{\label{gtbasistran}
E^{l,l} \ket{\bm\ld} &= (r_l^{\bm\ld}-r_{l-1}^{\bm\ld}) \ket {\bm\ld} \,, \\
E^{l,l+1} \ket{\bm\ld} &=\sum_{k=1}^l a_{k,l}^{\bm\ld}\ket{\bm\ld+1^{k,l}} \,, \\
E^{l+1,l} \ket{\bm\ld} &=\sum_{k=1}^l b_{k,l}^{\bm\ld}\ket{\bm\ld-1^{k,l}} \,,
}
where
\Als{\label{cslhwfgjw}
r_l^{\bm\ld} &= \sum\nolimits_{k=1}^l \ld_k^l \x{ and } r_0^{\bm\ld}=0 \,, \\
a_{k,l}^{\bm\ld} &= \bigg( -\f{ \prod_{i=1}^{l+1}(\hat \ld_i^{l+1}-\hat \ld_k^l) \prod_{i=1}^{l-1}(\hat \ld_i^{l-1}-\hat \ld_k^l-1) }{ \prod_{i=1,i\ne k}^l (\hat \ld_i^l-\hat \ld_k^l)(\hat \ld_i^l-\hat \ld_k^l-1) } \bigg)^{1/2} \,, \\
b_{k,l}^{\bm\ld} &= a_{k,l}^{\bm\ld-1^{k,l}} \,,
}
and $\hat \ld_i^l:=\ld_i^l-i$ for $i\le l$, $\hat \ld_i^l:=0$ for $i>l$, and zero factors in the products are skipped.
The coefficient $a_{k,l}^{\bm\ld}$ vanishes if $\bm\ld+1^{k,l}$ is not a valid pattern, and $b_{k,l}^{\bm\ld}$ vanishes if $\bm\ld-1^{k,l}$ is not a valid pattern.
It can be seen that the expressions~\eqref{gtbasistran} subsume~\eqref{clegjlhgerjo} as a special case of $\su(2)$ by observing that $\ket{j,m}=\big|\binom{2j\quad0}{j+m}\big\rangle$.

The Clebsch-Gordan transform in $\su(d)$ representation is introduced as follows.
The direct product of two irreps of $\su(d)$, $V_\ld$ and $V_\mu$,
is still a representation, and is in general reducible:
\Eq{\label{ch2:decomposs}
V_\ld \ot V_\mu =\Oplus_\nu  V_\nu^{\oplus N_{\ld,\mu}^\nu} \,,
}
where $N_{\ld,\mu}^\nu$ is the multiplicity of $V_\nu$ in the decomposition of $V_\ld\ot V_\mu$, or for short, the multiplicity of $\nu$ in $\ld\ot \mu$.
The multiplicity of $V_\nu$ can be determined by the Littlewood-Richardson rule.

The basis state of the space on the right-hand side of Eq.\ \eqref{ch2:decomposs} is written as
\Eq{\label{ciagao}
\ket{\bm\nu,\gm}=\sum_{\bm\ld,\bm\mu} C_{\bm\ld,\bm\mu}^{\bm\nu,\gm} \ket{\bm\ld}\ket{\bm\mu} \,,
}
where $\gm=1,\dots,N_{\ld,\mu}^\nu$ and $C_{\bm\ld,\bm\mu}^{\bm\nu,\gm}$ are Clebsch-Gordan coefficients (CGCs).

An algorithmic description of calculation of Clebsch-Gordan coefficients for $\fr{sl}(d)$ is given in~\cite{alex2011numerical}.
Denote the weight $w^{\bm\ld}:=(w_1^{\bm\ld},\dots,w_{d-1}^{\bm\ld})$ where $w_l^{\bm\ld} = r_l^{\bm\ld} - \f{1}{2}(r_{l+1}^{\bm\ld}+r_{l-1}^{\bm\ld})$.
Applying $J_{z,l}:=\f{1}{2}(E^{l,l}-E^{l+1,l+1})$ to both sides of~\eqref{ciagao} yields $w_l^{\bm\nu} \ket{\bm\nu,\gm}=\sum_{\bm\ld,\bm\mu} C_{\bm\ld,\bm\mu}^{\bm\nu,\gm} (w_l^{\bm\ld}+w_l^{\bm\mu}) \ket{\bm\ld}\ket{\bm\mu}$.
It follows that $C_{\bm\ld,\bm\mu}^{\bm\nu,\gm}\ne 0$ only if $w^{\bm\nu}=w^{\bm\ld}+w^{\bm\mu}$.

\appendix{ Technical derivation for Section~\ref{sec3}}\label{AppB}
In this Section we present proofs of Propositions \ref{ch2-prop-expec} and \ref{ch2-prop-higher}, in which Lemmas \ref{ch2:lem12}, \ref{lem-a2} and \ref{lem14} will be used.

Throughout we write $f\lesssim g$ for two positive functions $f$ and $g$ of $s$ if $f(s)\le cg(s)$ holds for some constant $c$ and any sufficiently large $s$,
write $f\simeq g$ if both $f\lesssim g$ and $g\lesssim f$ hold, and write $f\sim g$ if $f(n)/g(n)\to 1$ as $n\to\infty$.

\Lem{\label{ch2:lem12}
  Let $N(n,k)$ denote the multiplicity of irrep $V_{(k)}$ in $V_{(s)}^{\ot n}$ for $n\ge 2$.
  Then, as $s\to\infty$,
  \Eq{
  N(n,k)\lesssim s^{n-2} \x{ for each } k \,,
  }
  and
  \Eq{
  N(n,k)\simeq s^{n-2} \x{ for each } k \in[s, (n-1)s] \,.
  }
}
\proof{\!.
  Consider $V_{(s)}^{\ot n}=\Oplus_k V_{(k)}^{\oplus N(n,k)}$ for $s\in\f{1}{2}\Nb$.
  When $k>ns$ or $k<0$, set $N(n,k)=0$.
  Note that $V_{(j)}\ot V_{(s)}$ contains $V_{(k)}$ iff $|s-j|\le k\le s+j$.
  Thus $N(n,k)=N(n-1,k-s)+\cdots+N(n-1,k+s)$.
  It follows that by induction for $0\le k\le ns$, $N(n,k)\le (2s+1)\max_{k'}N(n-1,k')\le \cdots\le (2s+1)^{n-2}\max_{k''}N(2,k'')=(2s+1)^{n-2}$.
  Conversely, for $s\le k\le (n-1)s$, $N(n,k)\ge s\min_{s\le k'\le (n-2)s}N(n-1,k')\ge \cdots\ge s^{n-2} N(2,s)=s^{n-2}$.
}

\Lem{\label{lem-a2}
Let $j,\Dt,m$ be finite and fixed, and $m_1$ may depends on $j_1$.
As $j_1\to\infty$,
\Eq{\label{ch-s-cgc1}
|C^{j_1,j_1+\Dt,j}_{m_1,j-m_1,j}| \simeq j_1^{-j-\f{1}{2}}(j_1-m_1)^{\f{1}{2}(j+\Dt)}(j_1+m_1)^{\f{1}{2}(j-\Dt)} \,,
}
and
\Als{\label{ch-s-cgc2}
|C^{j_1,j_1+\Dt,j}_{m_1,m-m_1,m}| \lesssim j_1^{-j-\f{1}{2}} \sum_{k=0}^{j-\Dt} & (j_1-m_1)^{j-k+\f{1}{2}(m-\Dt)} \\
& \cdot(j_1+m_1)^{k+\f{1}{2}(\Dt-m)} \,.
}
}

\proof{\!.
Eq.\ \eqref{ch-s-cgc1} directly follows from Racah's formula.
Due to \eqref{racah}, $|C^{j_1,j_1+\Dt,j}_{m_1,m-m_1,m}|\le \sqrt{C_1C_2}C_3$, where 
\Aln{
C_1 =& \f{(2j+1)(j-\Dt)!(j+\Dt)!(2j_1+\Dt-j)!}{(2j_1+\Dt+j+1)!} \simeq j_1^{-2j-1} \,, \\
C_2 =& (j+m)!(j-m)!(j_1+m_1)!(j_1-m_1)!\\
& \cdot (j_1-m_1+\Dt+m)!(j_1+m_1+\Dt-m)! \,, \\
C_3 =& \sum\nolimits_{k=0}^{j-\Dt} f_k^{-1} \,,
}
with $f_k:=(j_1-m_1-j+\Dt+k)!(j_1+m_1-k)!(j-\Dt-k)!(j+m-k)!k!(\Dt-m+k)!$.
The sum in $C_3$ is extended over those $k$ such that the argument of every factorial is nonnegative (thus $0\le k\le j-\Dt$).
Note that $C_3 \le \sum_k ((j_1-m_1-j+\Dt+k)!(j_1+m_1-k)!)^{-1}$.
It follows that
\Aln{
\sqrt{C_2}C_3 &\lesssim \sum_k \sqrt{\gm_k} \\
&\simeq \sum_k(j_1-m_1)^{j-k+\f{1}{2}(m-\Dt)} (j_1+m_1)^{k+\f{1}{2}(\Dt-m)} \,,
}
where $\gm_k:=\f{(j_1+m_1)!(j_1-m_1)!(j_1-m_1+\Dt+m)!(j_1+m_1+\Dt-m)!}{(j_1-m_1-j+\Dt+k)!^2 (j_1+m_1-k)!^2}$, completing the proof of \eqref{ch-s-cgc2}.
}

\Lem{\label{lem14}
  For any nonnegative integers $t$ and $r$, $\sum_{i=1}^{n-1} i^t(n-i)^r \simeq n^{t+r+1} $ as $n\to\infty$.
}

\proof{\!.
  By Faulhaber's formula, $\sum_{k=1}^n k^t\sim \f{n^{t+1}}{t+1}$ as $n\to\infty$.
  The case $rt=0$ is readily verified, and we now prove the case $r,t>0$.
  We have
  \Aln{
  \sum_{i=1}^n i^t(n-i)^r &= \sum_{k=0}^r (-1)^k \binom{r}{k} n^{r-k} \sum_{i=1}^n i^{t+k}  \\
  & \sim  \sum_{k=0}^r (-1)^k \binom{r}{k} n^{r-k} \f{n^{t+k+1}}{t+k+1} \\
  & =: F(t+1,r)n^{t+r+1} \,,
  }
  where the second line holds if the coefficient $F(t+1,r)$ is nonzero.
  To complete the proof of this Lemma, it suffices to show $F(t+1,r)\ne 0$.
  We now prove that
  \Eq{\label{fnwjwt}
    F(t,r):=\sum_{k=0}^r (-1)^k \binom{r}{k} \f{1}{k+t} = \f{1}{t\binom{r+t}{t}}
  }
  for positive integers $t,r$.

  For any positive integer $m$, denote
  \Eq{\label{xx1}
  f(m):=\sum_{j=0}^r(-1)^j\binom{r+m}{j+m} \,.
  }
  By calculating $f(m)+f(m+1)$, we have $f(m)=\binom{r+m-1}{m-1}$.
  
  Now we calculate $F(1,r)$ and $F(2,r)$.
  It holds that $F(1,r)=\sum_{k=0}^r(-1)^k\binom{r}{k}\f{1}{k+1}=\f{1}{r+1}\sum_{k=0}^r(-1)^k\binom{r+1}{k+1}=\f{1}{r+1}$ and that $F(2,r)= \f{1}{r+1}\sum_{k=0}^r(-1)^k\binom{r+1}{k+1}- \f{1}{(r+1)(r+2)}\sum_{k=0}^r(-1)^k\binom{r+2}{k+2}
  =\f{1}{(r+1)(r+2)}$, where we have used Eq.\ \eqref{xx1}.

  Since $F(t,r+1)=\sum_{k=0}^{r+1}(-1)^k\binom{r+1}{k}\f{1}{k+t}$, we have
  \Aln{
  F & (t,r)-F(t,r+1) \\
  &= \sum_{k=1}^r (-1)^{k+1}\binom{r}{k-1}\f{1}{k+t} - (-1)^{r+1}\f{1}{r+1+t} \\
  &=\sum_{k=0}^{r-1} (-1)^k \binom{r}{k} \f{1}{k+t+1} +(-1)^r\f{1}{r+1+t} \\
  &=F(t+1,r)
  }

  Consequently, by induction on $t$, using the recurrence relation $F(t+1,r)=F(t,r)-F(t,r+1)$ and the expressions of $F(1,r)$ and $F(2,r)$, we 
  get~\eqref{fnwjwt}.
}

\proof{of Proposition~\ref{ch2-prop-expec}.
Since $\{\ket{ jmj_1j_2\alp_{j_1}\bt_{j_2} }\}$ is a basis of the near-invariant space,
\Aln{
\1_\x{inv}^{\ot 2}+W_\x{inv} = \sum \ 
& \big(
  \big| jmj_1j_2\alp_{j_1}\bt_{j_2} \big\rangle\big| j'm'j'_1j'_2\alp'_{j'_1}\bt'_{j'_2} \big\rangle \\
  &+ \big| j'm'j'_1j'_2\alp'_{j'_1}\bt'_{j'_2} \big\rangle\big| jmj_1j_2\alp_{j_1}\bt_{j_2} \big\rangle
  \big) \\
& \big( \big\langle jmj_1j_2\alp_{j_1}\bt_{j_2} \big|\big\langle j'm'j'_1j'_2\alp'_{j'_1}\bt'_{j'_2} \big|
  \big) \,,
}
the sum over $j,m,j_1,j_2,\alp_{j_1},\bt_{j_2},j',m',j'_1,j'_2,\alp'_{j'_1},\bt'_{j'_2}$.

Using Eq.\ \eqref{cgc3}, we have
\Aln{
  \tr\big( & W_\ty{AA'} (\ket{jmj_1j_2\alp_{j_1}\bt_{j_2} }\ket{ j'm'j'_1j'_2\alp'_{j'_1}\bt'_{j'_2}}) \\
  & \phantom{www} (\bra{jmj_1j_2\alp_{j_1}\bt_{j_2} }\bra{ j'm'j'_1j'_2\alp'_{j'_1}\bt'_{j'_2}}) \big) \\
  & = \sum_{m_1} \dt_{j_1,j'_1} \dt_{\alp_{j_1},\alp'_{j_1}} \big( C^{j_1,j_2,j}_{m_1,m-m_1,m} \big)^2    \big( C^{j_1,j'_2,j'}_{m_1,m'-m_1,m'} \big)^2 \,,
}
and similarly,
\Aln{
  \tr\big( & W_\ty{AA'} (\ket{j'm'j'_1j'_2\alp'_{j'_1}\bt'_{j'_2} }\ket{ jmj_1j_2\alp_{j_1}\bt_{j_2} }) \\
  &\phantom{www} (\bra{jmj_1j_2\alp_{j_1}\bt_{j_2} }\bra{ j'm'j'_1j'_2\alp'_{j'_1}\bt'_{j'_2}}) \big) \\
  & = \sum_{m_2} \dt_{j_2,j'_2} \dt_{\bt_{j_2},\bt'_{j_2}} \big( C^{j_1,j_2,j}_{m-m_2,m_2,m} \big)^2
  \big( C^{j'_1,j_2,j'}_{m'-m_2,m_2,m'} \big)^2 \,.
}

It follows that
\Eqn{
\tr(W_\ty{AA'} \1_\x{inv}^{\ot 2}) = \sum \big(C^{j_1,j_2,j}_{m_1,m-m_1,m}\big)^2 \big(C^{j_1,j'_2,j'}_{m_1,m'-m_1,m'}\big)^2 \,,
}
with the sum over $j,m,j_1,j_2,\alp_{j_1},\bt_{j_2},j',m',j'_2,\bt'_{j'_2},m_1$, and
\Eqn{
\tr(W_\ty{AA'} W_\x{inv}) = \sum \big(C^{j_1,j_2,j}_{m-m_2,m_2,m}\big)^2 \big(C^{j'_1,j_2,j'}_{m'-m_2,m_2,m'}\big)^2 \,,
}
with the sum over $j,m,j_1,j_2,\alp_{j_1},\bt_{j_2},j',m',j'_1,\alp'_{j'_1},m_2$.

Denoting $j_2:=j_1+\Dt$ and $j'_2:=j_1+\Dt'$, noticing $|\Dt|\le j$ and $|\Dt'|\le j'$, we have
\Aln{
  &\tr (W_\ty{AA'} \1_\x{inv}^{\ot 2}) \\
  & \ge     \sum
    \big(C^{j_1,j_1+\Dt,j}_{m_1,j-m_1,j}\big)^2 \big(C^{j_1,j_1+\Dt',j'}_{m_1,j'-m_1,j'}\big)^2 \\
  & \simeq  \sum
    j_1^{-2j-2j'-2} (j_1-m_1)^{j+j'+\Dt+\Dt'} (j_1+m_1)^{j+j'-\Dt-\Dt'} \\
  & \simeq  \sum_{j,j_1,\Dt,\alp_{j_1},\bt_{j_1+\Dt},j',\Dt',\bt'_{j_1+\Dt'}} j_1^{-1} \\
  & \gtrsim \sum_{j_1} N(p,j_1) N(q,j_1)^2 j_1^{-1} \\
  & \simeq  \sum_{j_1} s^{p-2}(s^{q-2})^2j_1^{-1} \\
  & \sim  s^{p+2q-6}\ln(p-1) \,,
}
where the sums in second and third lines are both over $j,j_1,\Dt,\alp_{j_1},\bt_{j_1+\Dt},j',\Dt',\bt'_{j_1+\Dt'}$ and $|m_1|< j_1$.
Thus $\tr(W_\ty{AA'} \1_\x{inv}^{\ot 2}) \gtrsim s^{p+2q-6}$.
In the above derivation, the first line uses the special cases $m=j$ and $m'=j'$, the second line uses Eq.\ \eqref{ch-s-cgc1}, the third line uses Lemma~\ref{lem14}, the fourth line with the sum over $j_1\in[s,(p-1)s]$ uses the special case $\Dt=\Dt'=0$, the fifth line uses Lemma~\ref{ch2:lem12}, and the last line uses the asymptotics of the harmonic series, i.e., $\sum_{k=1}^n \f{1}{k}\sim \ln n$ as $n\to\infty$.

In the other direction,
\Aln{
  & \tr(W_\ty{AA'} \1_\x{inv}^{\ot 2}) \\
  & \le     \sum
    j_1^{-2j-1} \sum_{k=0}^{j-\Dt} (j_1-m_1)^{2j-2k-\Dt+m} (j_1+m_1)^{2k+\Dt-m} \\
  &  \cdot j_1^{-2j'-1}  \sum_{k'=0}^{j'-\Dt'} (j_1-m_1)^{2j'-2k'-\Dt'+m'} (j_1+m_1)^{2k'+\Dt'-m'} \\
  & \simeq  \sum_{j,j_1,\Dt,\alp_{j_1},\bt_{j_1+\Dt},j',\Dt',\bt'_{j_1+\Dt'}} j_1^{-1} \\
  & \lesssim \max_{j_2} \sum_{j_1} N(p,j_1) N(q,j_2)^2 j_1^{-1} \\
  & \lesssim  s^{p+2q-6}\ln(ps) \,,
}
where the first line uses Eq.\ \eqref{ch-s-cgc2} and that $(\sum_{k=1}^na_k)^2\le n\sum_{k=1}^n a_k^2$, the unspecified sum in the second line is over $j,j_1,\Dt,\alp_{j_1},\bt_{j_1+\Dt},j',\Dt',\bt'_{j_1+\Dt'},m_1$, the `$\simeq$' is due to Lemma~\ref{lem14}, and the last line uses Lemma~\ref{ch2:lem12}.

Consequently,
\Eqn{
s^{p+2q-6} \lesssim \tr(W_\ty{AA'} \1_\x{inv}^{\ot 2}) \lesssim s^{p+2q-6}\ln s \,.
}

In the same way, we have
\Eqn{
s^{2p+q-6} \lesssim \tr(W_\ty{AA'} W_\x{inv}) \lesssim s^{2p+q-6}\ln s \,.
}

Since $\{\ket{ jmj_1j_2\alp_{j_1}\bt_{j_2} }\}$ is a basis of the near-invariant space, its dimension $d_\x{inv}$ is equal to
\Eq{\label{inv-su2-dimension}
  \sum_{j,m,\Dt}\sum_{j_1} N(p,j_1)N(q,j_1+\Dt) \simeq \sum_{j_1} s^{p-2}s^{q-2} \simeq s^{n-3} \,.
}

Thus,
\Aln{
\f{s^{p+2q-6}+s^{2p+q-6}}{(s^{n-3})^2} &\lesssim \E_\phv \tr(\phv_\ty{A}^2) \\
 & \lesssim \f{(s^{p+2q-6}+s^{2p+q-6})\ln s}{(s^{n-3})^2} \,,
}
completing the proof by noticing $p\le q$.
}

\proof{of Proposition~\ref{ch2-prop-higher}.
Using formula \eqref{twirl2},
\Eq{
\E (\tr\phv_\ty{A}^2)^2 =\f{1}{(d_\x{inv})^{\uparrow4}} \tr\big( (W_{(12)}^\ty{A} W_{(34)}^\ty{A})\sum\nolimits_{\pi\in S_4}W_\pi \big)
}
and
\Eq{\label{vweuhbfe}
(\E \tr\phv_\ty{A}^2)^2=\f{1}{d_\x{inv}^2(d_\x{inv}+1)^2} \tr\big( (W_{(12)}^\ty{A} W_{(34)}^\ty{A})\!\!\sum_{\pi\in S_2\times S_2}\!\!W_\pi \big) ,
}
where $(d_\x{inv})^{\uparrow 4}$ is the rising factorial, $W_{(12)}^\ty{A}$ swaps the first and second system $A$, and $W_{(34)}^\ty{A}$ swaps the third and fourth system $A$.
Thus
\Als{\label{goehgvjkjlwe}
&\f{\E(\tr\phv_\ty{A}^2)^2}{(\E\tr\phv_\ty{A}^2)^2}-1 = \f{-4d_\x{inv}-6}{(d_\x{inv}+2)(d_\x{inv}+3)} \\
&+ \f{(d_\x{inv}-1)!}{(d_\x{inv}+3)!} \f{1}{(\E\tr\phv_\ty{A}^2)^2} \sum_{\pi\in S_4\backslash S_2\times S_2} \tr \big( (W_{(12)}^\ty{A} W_{(34)}^\ty{A})W_\pi \big) \,.
}

Together with Eqs.\ \eqref{inv-su2-dimension} and~\eqref{vweuhbfe}, it follows that
\Eq{\label{fdk234}
\f{\E (\tr\phv_\ty{A}^2)^2 }{(\E\tr\phv_\ty{A}^2)^2}-1
\lesssim s^{-2p-4q+12} \!\!\!\!\sum_{\pi\in S_4\backslash S_2\times S_2}\!\!\! \tr \big( (W_{(12)}^\ty{A} W_{(34)}^\ty{A})W_\pi \big).
}

In the following, for $\pi\in S_4\backslash S_2\times S_2$, denote $t_i:=\pi^{-1}(i)$ for $i\in\{1,2,3,4\}$.
We have
\Aln{
  W_\pi = \sum 
  &\big| j^{t_1} m^{t_1} j_1^{t_1} j_2^{t_1} \alp_{j_1^{t_1}}^{t_1} \bt_{j_2^{t_1}}^{t_1} \big\rangle
  \big| j^{t_2} m^{t_2} j_1^{t_2} j_2^{t_2} \alp_{j_1^{t_2}}^{t_2} \bt_{j_2^{t_2}}^{t_2} \big\rangle \\
  &\big| j^{t_3} m^{t_3} j_1^{t_3} j_2^{t_3} \alp_{j_1^{t_3}}^{t_3} \bt_{j_2^{t_3}}^{t_3} \big\rangle
  \big| j^{t_4} m^{t_4} j_1^{t_4} j_2^{t_4} \alp_{j_1^{t_4}}^{t_4} \bt_{j_2^{t_4}}^{t_4} \big\rangle
  \\
  &\bra{ j^1 m^1 j_1^1 j_2^1 \alp_{j_1^1}^1 \bt_{j_2^1}^1 }
  \bra{ j^2 m^2 j_1^2 j_2^2 \alp_{j_1^2}^2 \bt_{j_2^2}^2 } \\
  &\bra{ j^3 m^3 j_1^3 j_2^3 \alp_{j_1^3}^3 \bt_{j_2^3}^3 }
  \bra{ j^4 m^4 j_1^4 j_2^4 \alp_{j_1^4}^4 \bt_{j_2^4}^4 } \,,
}
where the sum is over $j^k,m^k,j_1^k,j_2^k,\alp^k_{j_1^k},\bt^k_{j_2^k}$ for $k\in\{1,2,3,4\}$, and the superscripts indicate different systems.
In the rest of this proof, $k$ in each sum also ranges over $\{1,2,3,4\}$.

Using
\Aln{
\big| j^k m^k j_1^k j_2^k \alp_{j_1^k}^k \bt_{j_2^k}^k \big\rangle &= \sum_{} \big| j_1^km_1^k\alp_{j_1^k}^k \big\rangle \big| j_2^km_2^k\bt_{j_2^k}^k \big\rangle C^{j_1^kj_2^kj^k}_{m_1^km_2^km^k}
}
with the sum over $m_1^k,m_2^k$ such that $m_1^k+m_2^k=m^k$, and
\Aln{
\big\langle j^k m^k j_1^k j_2^k \alp_{j_1^k}^k \bt_{j_2^k}^k \big| &= \sum \big\langle j_1^k\hat m_1^k\alp_{j_1^k}^k \big| \big\langle j_2^k\hat m_2^k\bt_{j_2^k}^k \big| C^{j_1^kj_2^kj^k}_{\hat m_1^k\hat m_2^km^k} \,,
}
with the sum over $\hat m_1^k,\hat m_2^k$ such that $\hat m_1^k+\hat m_2^k=m^k$,
we have
\Aln{
  \tr \big( (W_{(12)}^\ty{A} W_{(34)}^\ty{A})W_\pi \big)
  = \sum_{k,j^k,m^k,j_1^k,j_2^k}  T_\x{cgc} T_{\alp\bt} \,,
}
where
\Aln{
  T_{\alp\bt}= \sum_{ k,\alp^k_{j_1^k},\bt^k_{j_2^k} } 1 \,,
}
and $T_\x{cgc}$ is equal to
\Aln{
  & \sum_{ k,m_1^k+m_2^k=m^k }
  C^{j_1^1,j_2^1,j^1}_{m_1^1,m_2^1,m^1}
  C^{j_1^2,j_2^2,j^2}_{m_1^2,m_2^2,m^2}
  C^{j_1^3,j_2^3,j^3}_{m_1^3,m_2^3,m^3}
  C^{j_1^4,j_2^4,j^4}_{m_1^4,m_2^4,m^4} \\
  & \phantom{wwwww}
  C^{j_1^{t_2},j_2^1,j^1}_{m_1^{t_2},m_2^{t_1},m^1}
  C^{j_1^{t_1},j_2^2,j^2}_{m_1^{t_1},m_2^{t_2},m^2}
  C^{j_1^{t_4},j_2^3,j^3}_{m_1^{t_4},m_2^{t_3},m^3}
  C^{j_1^{t_3},j_2^4,j^4}_{m_1^{t_3},m_2^{t_4},m^4} .
}

In this paper we use $\#\pi$ to denote the number of disjoint cycles in permutation $\pi$.
In the following, we consider $\pi\in S_4\backslash (S_2\times S_2)$, for which the pair $(\#\pi,\#((12)(34)\pi))$ belongs to the set $\{(3,1),(2,2),(1,3),(1,1)\}$.
Using Lemma~\ref{lem14} and Eq.\ \eqref{ch-s-cgc2}, we have $T_\x{cgc} \lesssim 1$ for each such $\pi$.

When $\#\pi=1\x{ or }2$,
\Aln{
\tr & (W^\ty{A}_{(12)}W^\ty{A}_{(34)}W_\pi) \\
& \lesssim\! \sum_{j_1,\Dt}\! N(p,j_1)^{\#((12)(34)\pi)}N(q,j_1+\Dt)^{\#\pi}.
}
Thus, $\tr(W^\ty{A}_{(12)}W^\ty{A}_{(34)}W_\pi) \lesssim s^{3p+q-7}$ for $\#\pi=1$, and $\tr(W^\ty{A}_{(12)}W^\ty{A}_{(34)}W_\pi) \lesssim s^{2p+2q-7}$ for $\#\pi=2$.

Specially, when $\#\pi=3$, we have $T_\x{cgc}\simeq (j_1^1)^{-2}$, thus $\sum_{k,j_1^k}T_\x{cgc}\lesssim 1$, hence
\Aln{
\tr\!\big(W^\ty{A}_{(12)}\!W^\ty{A}_{(34)}\!W_\pi\!\big)
&\!\lesssim\! \max_{j_1} N(p,j_1)^{\#((12)(34)\pi)}N(q,j_1\!\!+\!\!\Dt)^{\#\pi} \\
&\!\lesssim s^{p+3q-8} \,.
}

Consequently, using Eq.\ \eqref{fdk234}, it follows that
$$
\f{\E (\tr\phv_\ty{A}^2)^2 }{(\E\tr\phv_\ty{A}^2)^2}-1 \lesssim \max\{ s^{p-3q+5} , s^{-2q+5} , s^{-p-q+4} \}\,,$$
where the right-hand side vanishes as $s\to\infty$ provided $q\ge 3$ and $q\ge p\ge 2$.
}

\appendix{ Technical derivation for Section~\ref{sec4}}\label{AppC}

In this Section we first present the proof of Lemma~\ref{cnjksnvkvbad}, and then give the proof of Theorem~\ref{fwlngearnba} in which Lemmas \ref{Lemma9}, \ref{Lemma10} and \ref{Lemma11} will be used.

\proof{of Lemma \ref{cnjksnvkvbad}.
$V_\ld\ot V_\mu$ contains invariant state space iff $V_\ld\ot V_\mu$ contains $V_\nu$ for $\nu$ being $(\ld_1+\mu_d,\dots,\ld_1+\mu_d)$.
Due to the Littlewood-Richardson rule, we first add $\mu_d$ boxes filled with integer 1 in the first row, and then add $\mu_d$ boxes filled with integer 2 and $\ld_2-\ld_1$ boxes filled with integer 1.
Continuing so, we need to add $\mu_d+\ld_1-\ld_{d+1-k}$ boxes filled with $k$ for each $k$.
Thus $\ld_k+\mu_{d+1-k}=\ld_1+\mu_{d}$ for each $k$, that is, $\ld$ is dual to $\mu$.

Given a partition $\ld^d$, $V_{\ld^d}$ is an irrep of $\SU(d)$.
By the branching rule, as irrep of $\su(d-1)$, $V_{\ld^d}$ decomposes as $V_{\ld^d}=\Oplus_{\ld^{d-1}}V_{(\ld^d;\ld^{d-1})}$, where the direct sum is over all partitions $\ld^{d-1}$ interlacing $\ld^d$, and $V_{(\ld^d;\ld^{d-1})}:=\x{span}\{\ket{\bm\ld}:\bm\ld\in\x{GT}(\ld^d;\ld^{d-1})\}$.
The irrep $V_{\mu^d}$ decomposes in the same fashion.
It follows that
\Eq{
V_{\ld^d}\ot V_{\mu^d}=\Oplus_{\ld^{d-1},\mu^{d-1}} V_{(\ld^d;\ld^{d-1})}\ot V_{(\mu^d;\mu^{d-1})} \,.
}
Due to Littlewood-Richardson rule, in the orthogonal direct sum the term $V_{(\ld^d;\ld^{d-1})}\ot V_{(\mu^d;\mu^{d-1})}$ contains $V_{(0)}$ iff  $\ld^{d-1}$ and $\mu^{d-1}$ are dual to each other.
Continuing so, we have $C_{\bm\ld,\bm\mu}^{0}\ne 0$ iff $\bm\ld$ is dual to $\bm\mu$.
}

The CGC for dual pair of GT patterns is given as follows.
\Lem{\label{Lemma9}
  For each dual pair of partitions $\ld$ and $\mu$, and for each dual pair of GT patterns $\bm\ld\in\x{GT}(\ld)$ and $\bm\mu\in\x{GT}(\mu)$, $|C_{\bm\ld,\bm\mu}^0|= (\dim V_\ld)^{-1/2} $.
}

\proof{\!.
The irrep $V_{(0)}$ is contained in decomposition of $V_\ld\ot V_\mu$, hence
\Eq{\label{vkdhg}
  \ket0=\sum_{\bm\ld} C_{\bm\ld,\bm\mu}^0 \ket{\bm\ld}\ket{\bm\mu}
}
for $\ket 0\in V_{(0)}$.

For $l\in[d-1]$, applying $E^{l,l+1}$ to Eq.\ \eqref{vkdhg}, we have
\Aln{
  0 &= \sum_{k\in[l],\bm\ld} C_{\bm\ld,\bm\mu}^0 \big( a^{\bm\ld}_{k,l} \ket{\bm\ld+1^{k,l}} \ket{\bm\mu} + a^{\bm\mu}_{k,l} \ket{\bm\ld} \ket{\bm\mu+1^{k,l}} \big) \\
  &= \sum_{k\in[l],\bm\ld} C_{\bm\ld,\bm\mu}^0 \big( a^{\bm\ld}_{l+1-k,l} \ket{\bm\ld+1^{l+1-k,l}} \ket{\bm\mu} \\
  &\phantom{wwwwwww}+ a^{\bm\mu}_{k,l} \ket{\bm\ld} \ket{\bm\mu+1^{k,l}}  \big) \,.
}
Thus the coefficient of $\ket{\bm\ld} \ket{\bm\mu+1^{k,l}}$ vanishes:
\[
  C_{\bm\ld-1^{l+1-k,l},\bm\mu+1^{k,l}}^0 a^{\bm\ld-1^{l+1-k,l}}_{l+1-k,l} + C_{\bm\ld,\bm\mu}^0 a^{\bm\mu}_{k,l} =0 \,.
\]

It can be verified from Eq.\ \eqref{cslhwfgjw} that
$a_{l+1-k,l}^{\bm\ld-1^{l+1-k,l}} = a_{k,l}^{\bm\mu}
$.
It follows that $C_{\bm\ld-1^{l+1-k,l},\bm\mu+1^{k,l}}^0 + C_{\bm\ld,\bm\mu}^0 =0$ for each pair $(\bm\ld,\bm\mu)$ of dual patterns.
Thus for any dual pairs $(\bm\ld,\bm\mu)$ and $(\bm{\ld'},\bm{\mu'})$, either $C_{\bm\ld,\bm\mu}^0=C_{\bm{\ld'},\bm{\mu'}}^0$ or $C_{\bm\ld,\bm\mu}^0=-C_{\bm{\ld'},\bm{\mu'}}^0$.
Using the normalization condition of~\eqref{vkdhg} and the fact that $\x{GT}(\ld)$ has cardinality $\dim (V_\ld)$, the result follows.
}

\Lem{\label{Lemma10}
  Let $\mu=\mu^p=(\mu_1,\mu_2,\dots,\mu_{\min\{p,d\}})$ be any strictly decreasing sequence of positive numbers that sum to $p$.
  Then, as $s\to\infty$,

  (i) $N(\mu s)\simeq s^{(p-1)(p-2)/2}$ when $p\le d$,
  
  (ii) $N(\mu s)\simeq s^{(d-1)(d-2)/2+(d-1)(p-d)}$ when $p\ge d$,

  (iii) $N(\ld^p)\lesssim N(\mu s)$ for any $\ld^p\in\x{Par}(ps,\min\{p,d\})$.
}

\proof{\!.
  Without loss of generality we here consider the closest partition in $\x{Par}(ps,\min\{p,d\})$ to $\mu s$ if $\mu s$ itself is not a valid partition, since the distance between the two vectors vanishes as $s\to\infty$.

  For (i), let $\mu^{p-1}=(\mu^{p-1}_1,\dots,\mu^{p-1}_{p-1})$ be a strictly decreasing sequence of sum $p-1$.
  By Littlewood-Richardson rule, the decomposition of $\mu^{p-1}s\ot(s)$ contains $\mu s$ iff $\mu^{p-1}$ interlaces $\mu$, i.e., $\mu_1\ge \mu^{p-1}_1\ge \mu_2 \ge \mu^{p-1}_2\ge\cdots\ge\mu^{p-1}_{p-1}\ge\mu_p$.
  For given $\mu$, such a $\mu^{p-1}$ exists iff $\mu_1>1>\mu_p$.
  Continuing so, there exists a strictly decreasing sequence $\mu^{p-2}=(\mu^{p-2}_1,\dots,\mu^{p-2}_{p-2})$ such that $\mu^{p-2}s\ot(s)$ contains $\mu^{p-1} s$, iff $\mu_1+\mu_2>2>\mu_{p-1}+\mu_p$.
  Consequently, $\mu s$ has nonzero multiplicity in $(s)^{\ot p}$ iff $\sum_{i=1}^k\mu_i>k$ for each $k$.
  Indeed, the condition $\sum_{i=1}^k\mu_i>k$ holds for each $k$, since $\mu$ is a strictly decreasing sequence of sum $p$.
  Now we have obtained $\mu^k\in\Rb^k$ of sum $k$ for each $k\in\{2,\dots,p\}$, such that $\mu^k s$ is contained in $\mu^{k-1}s\ot (s)$ for each $k$.
  Since $N(\mu^2s)=1$, using induction it suffices to show $N(\mu^k s)\simeq s^{k-2}N(\mu^{k-1}s)$ for each $k$.

  Let $\tht=(\tht_1,\dots,\tht_k)$ be such that $\tht_1\in(0,1)$, $\tht_i\in\big(0,1-\sum_{j=1}^{i-1}\tht_j\big)$ for $2\le i\le k-1$, and $\tht_k=-\sum_{j=1}^{k-1}\tht_j$, and let $\tht'=(\tht'_1,\dots,\tht'_k)$ be defined similarly.
  For $\epsl>0$ small and fixed (independent of $s$), denote $\mu^k_\tht=\mu^k+\tht\epsl$.
  Set $\epsl$ so small that $\mu^{k-1}_\tht$ interlaces $\mu^k_{\tht'}$ for each $\tht,\tht'$.
  Since $\mu^{k-1}_\tht s$ has $\simeq s^{k-2}$ choices as $\tht$ varies for given $\mu^{k-1}$, and $N(\mu^k_{\tht'}s)\ge \sum_\tht N(\mu^{k-1}_\tht s)$, we have $N(\mu^k s)\gtrsim s^{k-2}N(\mu^{k-1}s)$.
  Due to the upcoming proof of (iii) (for $p\le d$), it holds that $N(\mu^k s)\lesssim s^{k-2}N(\mu^{k-1}s)$, completing the proof of (i).

  We now prove (iii) for the case $p\le d$.
  Since $N(\ld^2)=1$ for $\ld^2\in\x{Par}(2s,2)$, it suffices to show that $N(\ld^k)\lesssim s^{k-2} \max_{\ld^{k-1}} N(\ld^{k-1})$ for any $2\le k\le p$ and $\ld^k\in\x{Par}(ks,k)$, where the maximum is over $\ld^{k-1}\in\x{Par}((k-1)s,k-1)$.
  $\ld^k$ is contained in decomposition of $\ld^{k-1}\ot(s)$ for some $\ld^{k-1}$ iff $\ld^{k-1}$ interlaces $\ld^k$.
  Since $\ld^k_i-\ld^k_{i-1}\lesssim s$ for each $i$, and $\ld^{k-1}$ has sum $(k-1)s$, the number of $\ld^{k-1}$'s that interlace $\ld^k$ is $\lesssim s^{k-2}$.
  Thus $N(\ld^k)\lesssim s^{k-2} \max_{\ld^{k-1}} N(\ld^{k-1})$.

  The case $p>d$ in (ii) and (iii) is proved in the same way.
  For $p\ge d+1$, $N(\mu s)\simeq s^{d-1}N(\mu^{p-1}s) \simeq s^{(d-1)(p-d)}N(\mu^ds)\simeq s^{(d-1)(d-2)/2+(d-1)(p-d)}$.
}

\Lem{\label{Lemma11}
  Let $p\le q$, and denote $b:=\f{p+q}{d}$ and $t:=p+q-d$.
  Let $\mu=\mu^p=(\mu_1,\mu_2,\dots,\mu_{\min\{p,d\}})$, where $\mu_1<b$, be any strictly decreasing sequence of positive numbers that sum to $p$.
  If $t\le 0$, $N(\ld_*)=0$ for any $\ld\in\x{Par}(ps,\min\{p,d\})$.
  If $t>0$,

  (i) $N((\mu s)_*)\simeq s^{(d-1)(d-2)/2+(d-1)(q-d)}$ when $p\ge d$,
  
  (ii) $N((\mu s)_*)\simeq s^{(d-1)(d-2)/2+(d-1)(t-d)+(p+d-1)(d-p)/2}$ when $p\le d$ and $t\ge d$,
  
  (iii) $N((\mu s)_*)\simeq s^{(p+t-1)(q-d)/2+(t-1)(d-t)+(t-1)(t-2)/2}$ when $p\le d$, $t\le d$ and $q\ge d$,

  (iv) $N((\mu s)_*)\simeq s^{ (p-1)(d-p)+(t-1)(t-2)/2 }$ when $p\le d$, $t\le d$ and $q\le d$,

  (v) $N((\ld^p)_*)\lesssim N((\mu s)_*)$ for any $\ld^p\in\x{Par}(ps,\min\{p,d\})$.
}
\proof{\!.
  Notice that $\ld _*\in\x{Par}(qs)$ for $\ld \in\x{Par}(ps)$.  
  Since the proof idea is similar to that of Lemma~\ref{Lemma10}, a brief calculation is given as follows.

  For (i), when $p\ge d$, for any $\ld=(\mu_1,\dots,\mu_d)s\in\x{Par}(ps,d)$, $\ld_*=(b-\mu_d,\dots,b-\mu_1)s$ is a strictly decreasing sequence of sum $q$.
  Since $q\ge d$, and $\mu_1<b$, by Lemma~\ref{Lemma10}, $N(\ld_*)\simeq s^{(d-1)(d-2)/2+(d-1)(q-d)}$.

  When $p<d$, for any $\ld=\mu^ps=(\mu_1,\dots,\mu_p)s$ with $\mu_1< b$, we have $\ld_*=\nu^q s$, where $\nu^q:=(b,\dots,b,b-\mu_p,\dots,b-\mu_1)$ has sum $q$.
  Let $\nu^q$ be interlaced by $\nu^{q-1}$, $\nu^{q-1}$ be interlaced by $\nu^{q-2}$, and so on until $\nu^2$ be interlaced by $\nu^1=(1)$, where $\nu^k$ has sum $k$ for each $k$.
  Three subcases are calculated as follows.

  For (ii), when $t\ge d$, $\nu^qs$ has multiplicity $N(\nu^qs)\simeq s^p N(\nu^{q-1}s) \simeq s^{p+(p+1)+\cdots+(d-1)} N(\nu^ts) = s^{(p+d-1)(d-p)/2}N(\nu^ts)$, $\nu^ts$ has multiplicity $N(\nu^t s)\simeq s^{(d-1)(p+q-2d)}N(\nu^ds)$, and $\nu^ds$ has multiplicity $N(\nu^ds)\simeq s^{(d-1)(d-2)/2}$.
  Thus $N((\mu s)_*)=N(\nu^qs)$ is obtained.
  
  For (iii), when $t< d$ and $d\le q$, $\nu^qs$ has multiplicity $N(\nu^qs)\simeq s^p N(\nu^{q-1}s) \\ 
  \simeq s^{p+(p+1)+\cdots+(p+q-d-1)} N(\nu^d s)  = s^{(2p+q-d-1)(q-d)/2}N(\nu^ds)$, $\nu^ds$ has multiplicity $N(\nu^ds)\simeq s^{(p+q-d-1)(2d-p-q)} N(\nu^ts)$, and $\nu^ts$ has multiplicity $N(\nu^ts)\simeq s^{(p+q-d-1)(p+q-d-2)/2}$.
  Thus $N(\nu^qs)$ is obtained.
  
  For (iv), when $t< d$ and $d\ge q$, $\nu^qs$ has multiplicity $N(\nu^qs)\simeq s^{(p-1)(d-p)} N(\nu^t s)$, and $\nu^ts$ has multiplicity $N(\nu^ts)\simeq s^{(t-1)(t-2)/2}$.
}

\Lem{\label{josdhbdfhbkvkjsadn}
  For random invariant state $\phv_\ty{AB}$ as in Theorem 4, as $s\to\infty$,
  $$
  \E\tr\phv_\ty{A}^2 \simeq s^{-p(d-1)} \,,
  $$
  and
  $$
  \f{\E(\tr\phv_\ty{A}^2)^2}{(\E\tr\phv_\ty{A}^2)^2}-1 
  \lesssim  s^{(d-1)(-p-q+d+2)} \,.
  $$
}
\proof{\!.
  For $p\ge d$, $\ld$ has $s^{d-1}$ choices.
  By Weyl dimension formula, $\dim V_\ld\simeq s^{d(d-1)/2}$.
  Let $\mu=\mu^p=(\mu_1,\mu_2,\dots,\mu_h)$, where $\mu_1<b$, be any strictly decreasing sequence of positive numbers that sum to $p$, and let $\ld=\mu s$.
  Then by Lemmas~\ref{Lemma10} and~\ref{Lemma11},
  \Als{\label{fdkjlhfgwe}
  d_\x{inv}\!\! &= \sum_{\ld\vdash ps,\ld_*\vdash qs} N(\ld)N(\ld_*) \\
  &\simeq s^{d-1}s^{(d-1)(d-2)/2+(d-1)(p-d)}s^{(d-1)(d-2)/2+(d-1)(q-d)} \\
  &= s^{(d-1)(t-1)} \,,
  }
  and
  \Aln{
  \tr &(W_\ty{AA'}\1_\x{inv}^{\ot 2}) + \tr(W_\ty{AA'}W_\x{inv}) \\
  &\simeq s^{d-1}s^{(d-1)(d-2)/2+(d-1)(p-d)} \\
  &\quad \cdot(s^{(d-1)(d-2)/2+(d-1)(q-d)})^2 s^{-d(d-1)/2} \\
  &=s^{(d-1)(p+2q-2d-2)} \,,
  }
  from which 
  it follows that $\E\tr\phv_\ty{A}^2 \simeq s^{-p(d-1)}$.

  We now give a bound on the variance of $\tr\phv_\ty{A}^2$. 
  Using Eq.\ \eqref{fdkjlhfgwe}, and Eq.\ \eqref{goehgvjkjlwe} which still holds for the case $d\ge 3$, we have
  \Als{\label{cklwehfweflbfv}
  &\f{\E(\tr\phv_\ty{A}^2)^2}{(\E\tr\phv_\ty{A}^2)^2} -1 \\
  &\lesssim s^{-2(d-1)(p+2q-2d-2)} \sum_{\pi\in S_4\backslash S_2\times S_2} \tr \big( (W_{(12)}^\ty{A} W_{(34)}^\ty{A})W_\pi \big) .
  }

We have
  \Als{
  W_\pi 
  = \sum &
  \ket{ \bm\ld^{t_1} \alp^{t_1}_{\ld^{t_1}} }  \ket{ \bm\mu^{t_1} \bt^{t_1}_{\mu^{t_1}} }
  \ket{ \bm\ld^{t_2} \alp^{t_2}_{\ld^{t_2}} }  \ket{ \bm\mu^{t_2} \bt^{t_2}_{\mu^{t_2}} } \\
  &
  \ket{ \bm\ld^{t_3} \alp^{t_3}_{\ld^{t_3}} }  \ket{ \bm\mu^{t_3} \bt^{t_3}_{\mu^{t_3}} }
  \ket{ \bm\ld^{t_4} \alp^{t_4}_{\ld^{t_4}} }  \ket{ \bm\mu^{t_4} \bt^{t_4}_{\mu^{t_4}} } \\
  &
  \bra{ \bm\ld^{t_2} \alp^1_{\ld^1} }  \bra{ \bm\mu^{t_1} \bt^1_{\mu^1} }
  \bra{ \bm\ld^{t_1} \alp^2_{\ld^2} }  \bra{ \bm\mu^{t_2} \bt^2_{\mu^2} } \\
  &
  \bra{ \bm\ld^{t_4} \alp^3_{\ld^3} }  \bra{ \bm\mu^{t_3} \bt^3_{\mu^3} }
  \bra{ \bm\ld^{t_3} \alp^4_{\ld^4} }  \bra{ \bm\mu^{t_4} \bt^4_{\mu^4} } \\
  &
  \cdot C^0_{\bm\ld^{t_1},\bm\mu^{t_1}}
  C^0_{\bm\ld^{t_2},\bm\mu^{t_2}}
  C^0_{\bm\ld^{t_3},\bm\mu^{t_3}}
  C^0_{\bm\ld^{t_4},\bm\mu^{t_4}} \\
  &
  \cdot C^0_{\bm\ld^{t_2},\bm\mu^{t_1}}
  C^0_{\bm\ld^{t_1},\bm\mu^{t_2}}
  C^0_{\bm\ld^{t_4},\bm\mu^{t_3}}
  C^0_{\bm\ld^{t_3},\bm\mu^{t_4}} \\
  &+\x{terms irrelevant to } W^\ty{A}_{(12)}W^\ty{A}_{(34)}
  }
  where the sum is over $k,\ld^k,\alp^k_{\ld^k},\bt^k_{\mu^k},\bm\ld^k$, $\bm\ld^k,\bm{\hat\ld}^k\in\x{GT}(\ld^k)$, and $\bm\mu^k,\bm{\hat\mu}^k\in\x{GT}(\mu^k)$.

  In order to calculate $\tr(W^\ty{A}_{(12)}W^\ty{A}_{(34)}W_\pi)$, assume without loss of generality that
  $\bm\ld^{t_1}=\bm{\hat\ld}^2$,
  $\bm\ld^{t_2}=\bm{\hat\ld}^1$,
  $\bm\ld^{t_3}=\bm{\hat\ld}^4$,
  $\bm\ld^{t_4}=\bm{\hat\ld}^3$,
  $\bm\mu^{t_1}=\bm{\hat\mu}^1$,
  $\bm\mu^{t_2}=\bm{\hat\mu}^2$,
  $\bm\mu^{t_3}=\bm{\hat\mu}^3$,
  $\bm\mu^{t_4}=\bm{\hat\mu}^4$,
  $\alp^1_{\ld^1}=\alp^{t_2}_{\ld^{t_2}}$,
  $\alp^2_{\ld^2}=\alp^{t_1}_{\ld^{t_1}}$,
  $\alp^3_{\ld^3}=\alp^{t_4}_{\ld^{t_4}}$,
  $\alp^4_{\ld^4}=\alp^{t_3}_{\ld^{t_3}}$,
  $\bt^1_{\mu^1}=\bt^{t_1}_{\mu^{t_1}}$,
  $\bt^2_{\mu^2}=\bt^{t_2}_{\mu^{t_2}}$,
  $\bt^3_{\mu^3}=\bt^{t_3}_{\mu^{t_3}}$,
  $\bt^4_{\mu^4}=\bt^{t_4}_{\mu^{t_4}}$.

Thus
\Als{
  \tr(W^\ty{A}_{(12)}W^\ty{A}_{(34)}W_\pi) = \sum_{\ld^1,\ld^2,\ld^3,\ld^4} T_\x{cgc}T_{\alp\bt} \,,
}
where
\Aln{
  T_{\alp\bt} = \sum_{k,\alp^k_{\ld^k},\bt^k_{\mu^k}} 1 \simeq N(\ld)^{\#(W\pi)}N(\ld_*)^{\#\pi} \,,
}
and
\Als{\label{fdkjskjfcgc}
  T_\x{cgc} = \sum_{\bm\ld^1,\bm\ld^2,\bm\ld^3,\bm\ld^4}
  &
  C^0_{\bm\ld^{t_1},\bm\mu^{t_1}}
  C^0_{\bm\ld^{t_2},\bm\mu^{t_2}}
  C^0_{\bm\ld^{t_3},\bm\mu^{t_3}}
  C^0_{\bm\ld^{t_4},\bm\mu^{t_4}} \\
  &
  C^0_{\bm\ld^{t_2},\bm\mu^{t_1}}
  C^0_{\bm\ld^{t_1},\bm\mu^{t_2}}
  C^0_{\bm\ld^{t_4},\bm\mu^{t_3}}
  C^0_{\bm\ld^{t_3},\bm\mu^{t_4}}
}

Since the summand on the right-hand side of Eq.\ \eqref{fdkjskjfcgc} is nonzero only if $\bm\ld^{t_1}=\bm\ld^{t_2}$ and $\bm\ld^{t_3}=\bm\ld^{t_4}$.
It follows that $T_\x{cgc}\ne 0$ only if $\ld^{t_1}=\ld^{t_2}$ and $\ld^{t_3}=\ld^{t_4}$, in which case,
\Aln{
  T_\x{cgc} &= \sum_{\bm\ld^{t_1},\bm\ld^{t_3}} (\dim V_{\ld^{t_1}})^{-2}(\dim V_{\ld^{t_3}})^{-2}\\
  &= (\dim V_{\ld^{t_1}})^{-1}(\dim V_{\ld^{t_3}})^{-1} \\
  & \simeq s^{-d(d-1)} \,.
}

Thus
\Eqn{
  \tr(W^\ty{A}_{(12)}W^\ty{A}_{(34)}W_\pi) \simeq \sum_{\ld^{t_1},\ld^{t_3}} N(\ld)^{\#(W\pi)}N(\ld_*)^{\#\pi} s^{-d(d-1)} \,.
}

When $\#(W\pi)=1$ and $\#\pi=3$,
\Aln{
  \tr(W^\ty{A}_{(12)}W^\ty{A}_{(34)}W_\pi) &\simeq s^{2(d-1)} s^{(d-1)(d-2)/2+(d-1)(p-d)} \\
  & \phantom{w} \cdot (s^{(d-1)(d-2)/2+(d-1)(q-d)})^3 s^{-d(d-1)} \\
  &= s^{(d-1)(p+3q-3d-2)} \,.
}

Using Eq.\ \eqref{cklwehfweflbfv}, we have
\Eq{\label{cwjlhdeolf}
  \f{\E(\tr\phv_\ty{A}^2)^2}{(\E\tr\phv_\ty{A}^2)^2}-1 
  \lesssim s^{(d-1)(-p-q+d+2)} \,,
}
which vanishes as $s\to\infty$ if $p+q>d+2$.
Under the condition $q\ge p\ge d$, $p+q>d+2$ iff $q\ge 3$.

When $(\#\pi,\#((12)(34)\pi))$ equals $(2,2)$, $(1,3)$, or $(1,1)$, the estimate similar to Eq.\ \eqref{cwjlhdeolf} can be obtained.
Combining these estimates, we have 
\Aln{
  \f{\E(\tr\phv_\ty{A}^2)^2}{(\E\tr\phv_\ty{A}^2)^2}-1 
  &\lesssim \max\big\{ s^{(d-1)(-p-q+d+2)}, s^{(d-1)(-2q+d+2)}, \\
  & \phantom{www} s^{(d-1)(p-3q+d+2)}, s^{(d-1)(-p-3q+2d+4)} \big\} \\
  &= s^{(d-1)(-p-q+d+2)} \,.
}
}

\end{document}